\RequirePackage{fix-cm}
\documentclass[twocolumn,epjc3]{svjour3}  
\smartqed  % flush right qed marks, e.g. at end of proof
\RequirePackage{graphicx}
\usepackage{makecell}
\usepackage{multirow}
\usepackage{amsmath,amssymb,amsfonts}
\usepackage{mathrsfs}
\usepackage[title]{appendix}
\usepackage{xcolor}
\usepackage{textcomp}
\usepackage{manyfoot}
\usepackage{booktabs}
\usepackage{threeparttable}
\usepackage{algorithm}
\usepackage{algorithmicx}
\usepackage{algpseudocode}
\usepackage{listings}
\usepackage{pdflscape}
\usepackage{pgfplots}
\pgfplotsset{compat=1.18}
\usepackage{comment}
\journalname{}
\begin{document}

\title{A data-driven security quantification framework for IoT-based systems}%Vulnerability-Centric Quantitative Cybersecurity Risk Assessment for Enhanced IoT Dependability


\author{Alhassan Abdulhamid\thanksref{e1}
        \and
        Sohag Kabir\thanksref{e2}
        \and Ibrahim Ghafir\thanksref{e3} \and
        Ci Lei\thanksref{e4}
        %\and Omar Albahbouh %Aldabbas\thanksref{e5,addr2}%etc.
}

\thankstext{e1}{e-mail: a.abdulhamid2@bradford.ac.uk}
\thankstext{e2}{e-mail: s.kabir2@bradford.ac.uk}
\thankstext{e3}{e-mail: i.ghafir@bradford.ac.uk }
\thankstext{e4}{e-mail: c.lei1@bradford.ac.uk}
%\thankstext{e5}{e-mail:o.aldabbas@bau.edu.jo}


\institute{School of Computing and Engineering, University of Bradford, Bradford, United Kingdom %\label{addr1}
           %\and
           %Faculty of Engineering, Al-Balqa Applied University, Salt, Jordan \label{addr2}
}

\date{Received: date / Accepted: date}

\maketitle

\begin{abstract}
The Internet of Things (IoT) is integral to modern cyber-physical systems. Quantitative cybersecurity assessment in IoT environments remains challenging due to heterogeneous system architectures, evolving threat landscapes, and the limited availability of reliable probabilistic exploitability data. Although Attack Tree Analysis (ATA) provides a structured framework for modelling potential attack paths leading to system compromise, conventional ATA quantification often relies on subjective expert judgement or heuristic scoring schemes, which can introduce uncertainty and reduce analytical reproducibility. This study introduces a data-driven probabilistic security framework for IoT-based safety-critical systems by integrating Model-Based Systems Engineering (MBSE), ATA, and empirical vulnerability data. In the proposed framework, SysML models capture system architecture, from which attack trees are derived. Vulnerabilities are mapped as Basic Attack Steps and assigned exploitation probabilities using the Exploit Prediction Scoring System (EPSS). The attack tree is then represented as a Bayesian Network, enabling probabilistic reasoning, diagnostic inference, and vulnerability criticality analysis. The framework quantifies system compromise probabilities, identifies likely causes of attacks, and prioritises mitigation strategies. By combining architecture-driven modelling with real-world vulnerability intelligence, it provides a rigorous, reproducible approach for cybersecurity risk assessment in complex IoT environments.

%The Internet of Things (IoT) has become an integral part of modern life, making the security of these systems a critical concern. Attack Tree Analysis (ATA) is a widely used model for identifying security vulnerabilities in system architectures. However, assigning reliable probabilistic values to cybersecurity metrics within this artefact is often challenging due to limited vulnerability data, the evolving nature of cyber threats, and the complexity of IoT environments. Expert judgement, while valuable, can introduce subjectivity and uncertainty, limiting its effectiveness for precise cybersecurity quantification and IoT standardisation. This study introduces an enhanced ATA-based approach that integrates publicly available vulnerability exploitation data to improve cybersecurity assessment. By systematically mapping vulnerabilities in IoT architectures, assigning objective probabilistic scores, and linking them to attack objectives within the ATA framework, this method quantifies the probability of system failure due to cyber threats. The proposed approach evaluates the likelihood of confidentiality, availability, and data integrity breaches by leveraging real-world vulnerability data and logical dependencies within the attack tree. By reducing subjectivity and strengthening the analytical quantification of cyber attack impacts on system functionality, the proposed methodology enables a more rigorous and representative risk assessment framework, thereby supporting the development of robust and dependable IoT-enabled systems.
\keywords{Internet of Things \and Security Quantification \and Attack Tree \and Exploit Prediction Scoring System}
\end{abstract}
\section{Introduction}\label{sec1}

The Internet of Things (IoT) has transformed interactions with the physical world by embedding computational capabilities into everyday objects, enabling communication and coordination over the Internet. Through the integration of computation, communication, and mechatronic technologies, IoT supports real-time monitoring and control of physical systems and processes \cite{abdulhamid2022dependability,lefoane2023sequential}. This convergence enables intelligent, interconnected systems that enhance automation, responsiveness, and operational efficiency \cite{musa2023open,ayaz2019internet}. By bridging digital and physical domains, IoT facilitates sensing, analysis, and control across diverse application areas, including smart homes, industrial automation, healthcare, transportation, agriculture, and defence \cite{qureshi2020iotfc,raza2017iot,gope2020secure,boddu2023iot,lefoane2021machine}.

Despite these transformative capabilities, the increasing interconnectivity and heterogeneity of IoT systems substantially expand their attack surface. The integration of distributed sensing, wireless communication, edge intelligence, and cloud services introduces complex cyber-physical dependencies that are highly susceptible to cyber threats \cite{xing2020,qureshi2015framework,ghafir2016survey}. As IoT deployments continue to scale in critical infrastructures, cybersecurity becomes a fundamental determinant of system dependability.

Dependability in IoT systems refers to the ability of a system to deliver its intended service reliably and securely within a specified timeframe \cite{abdulhamid2022dependability}. According to \cite{avizienis2001fundamental}, dependability encompasses availability, reliability, integrity, confidentiality, safety, and maintainability. From a cybersecurity perspective, the confidentiality, integrity, and availability (CIA) triad forms the cornerstone of secure system operation. Availability ensures operational continuity, integrity guarantees correctness and consistency of system data, and confidentiality protects sensitive information from unauthorised access \cite{avizienis2001fundamental}.

Although significant research has addressed safety and reliability modelling in IoT systems \cite{xing2020,steiner2015qualitative,LGEN21,kabir2015reliability,abdulhamid2023overview,abdulhamid2025b,abdulhamid2024enhsaf,abdulhamid2023adaptation,rahman2023qualitative,abdulhamid2024reliability}, comparatively less attention has been devoted to rigorous quantitative modelling of cybersecurity attributes within unified dependability frameworks. Given the hybrid cyber-physical nature of IoT systems, cybersecurity analysis must be systematically integrated with safety and reliability considerations to achieve holistic dependability assessment.

IoT systems operate through a complex combination of wireless communication, intelligent processing, and cloud computing technologies \cite{abdulhamiddev,qureshi2015framework}. While these technologies enable seamless distributed functionality, they also increase exposure to cyber attacks such as remote exploitation, malware propagation, and denial-of-service attacks \cite{ghafir2015dns}. The constrained computational capabilities of many IoT devices, coupled with large-scale distributed deployment, further complicate the implementation of robust security mechanisms \cite{qureshi2015framework,ghafir2016survey,abdulhamid2023overview}.

Although security is commonly treated as a non-functional property (NFP), incorporating quantifiable cybersecurity assessment during the conceptual design phase remains challenging due to system heterogeneity, multiple attack vectors, and limited device resources \cite{xing2020,lefoane2022unsupervised}. Traditional vulnerability assessment approaches frequently rely on qualitative judgement or ordinal scoring schemes, limiting analytical rigour and reproducibility. While proactive security verification during system development is encouraged \cite{jacobs2021exploit,xing2020}, the absence of objective probabilistic quantification restricts informed risk-based decision-making.

Attack Tree Analysis (ATA) provides a structured framework for modelling potential attack paths leading to system compromise \cite{45,109}. By decomposing a top event (TE), typically representing a system breach, into logical combinations of attack goals and sub-goals, ATA enables estimation of system-level security risk \cite{109}. However, conventional ATA quantification often depends on subjective, cost-based, or expert-derived estimates \cite{45,109,abdulhamid2023overview}, resulting in uncertainty and limited standardisation. Unlike reliability engineering, where empirical failure data are often available, cybersecurity modelling has historically lacked consistent probabilistic exploitability data, leading to variability in risk estimation.

Recent advances in vulnerability intelligence have created opportunities for data-driven cybersecurity quantification. Several publicly available vulnerability databases support systematic analysis across cybersecurity domains \cite{jacobs2021exploit}. Among these, the Exploit Prediction Scoring System (EPSS) provides probabilistic estimates of real-world vulnerability exploitation \cite{janiszewski2022creating}. EPSS predicts the likelihood that a disclosed vulnerability will be exploited in the wild, generating probabilistic scores ranging from 0\% to 100\%. The framework leverages Common Vulnerabilities and Exposures (CVE) data, empirical exploit evidence, and community intelligence to compute dynamic exploitability metrics \cite{janiszewski2022creating,ali2011new,scarfone2009analysis,jacobs2021exploit}. By incorporating factors such as attack complexity, exploit availability, and prevailing threat conditions \cite{scarfone2009analysis}, EPSS provides an evidence-based measure of vulnerability exploitation probability.

Motivated by the need for rigorous and architecture-aware cybersecurity quantification, this study proposes a data-driven probabilistic security assessment framework for IoT-based safety-critical systems. The framework integrates Model-Based Systems Engineering (MBSE), attack tree modelling, and empirical exploitability data to quantify the likelihood of system compromise. System architecture is first represented using SysML models, from which attack trees are derived to capture potential attack propagation paths. Identified vulnerabilities are mapped to system components and incorporated as Basic Attack Steps (BAS), with exploitability probabilities assigned using EPSS data. This integration enables objective estimation of the probability of a TE representing system compromise.

To further enhance analytical capability, the attack tree structure is mapped to a Bayesian Network representation, enabling probabilistic reasoning beyond deterministic attack tree evaluation. This probabilistic formulation supports diagnostic inference and vulnerability criticality analysis, allowing the identification of vulnerabilities that most significantly contribute to system compromise and enabling prioritisation of mitigation strategies.

The proposed framework is domain-agnostic; however, smart agriculture is employed as a representative cyber-physical IoT case study to validate the methodology. In smart agriculture, IoT-based technologies and Unmanned Aerial Vehicles (UAVs) optimise irrigation, pest detection, crop monitoring, and resource management \cite{kumar2020smart}. Integrated sensor networks monitor soil moisture, temperature, humidity, and water levels \cite{dahane2020iot,gutierrez2013automated}, while deep learning techniques enhance pest and disease detection \cite{saranya2023comparative}. UAV-enabled systems support soil analysis and crop damage assessment \cite{maslekar2020application}, vertical farming utilises IoT and AI for resource efficiency \cite{kalantari2018opportunities}, and precision livestock farming leverages RFID, GPS, and smart geofencing \cite{mishra2023advanced}. The operational complexity and cyber-physical interdependencies of such systems make them suitable for validating quantitative cybersecurity assessment approaches.

To the best of our knowledge, limited research has combined architecture-driven modelling, empirical vulnerability exploitability data, and probabilistic reasoning within a unified framework for quantifying cybersecurity risk in IoT systems.

The specific contributions of this article are as follows:
\begin{itemize}
\item Development of a data-driven probabilistic framework for cybersecurity risk quantification in IoT-based safety-critical systems, integrating attack tree modelling with vulnerability intelligence within an MBSE context.

\item Integration of EPSS exploitability probabilities into attack tree structures to enable objective estimation of system-level compromise likelihood based on real-world vulnerability data.

\item Transformation of the attack tree representation into a Bayesian Network to support probabilistic reasoning, diagnostic inference, and vulnerability criticality analysis for prioritising cybersecurity mitigation strategies.

\item Demonstration of the proposed framework through a smart agriculture IoT case study, illustrating how vulnerability exploitation can propagate to system-level failure in safety-critical cyber-physical environments.
\end{itemize}

%The specific contributions of this article are as follows:
%\begin{itemize}
%\item Development of a data-driven probabilistic security quantification framework that integrates MBSE, attack tree modelling, and real-world vulnerability exploitability data.
%\item Incorporation of EPSS-derived exploitation probabilities into attack tree structures to objectively estimate system-level compromise likelihood.
%\item Mapping of attack tree models to Bayesian Networks to enable probabilistic inference and vulnerability criticality analysis for prioritising mitigation strategies.
%\item Demonstration of the proposed framework through an IoT-based smart agriculture case study representing a safety-critical cyber-physical system.
%\end{itemize}

The remainder of this paper is organised as follows. Section \ref{sec2} reviews the relevant background and related work. Section \ref{sec3} presents the proposed vulnerability-centric quantitative modelling approach, and Section \ref{casestudy} demonstrates its application through a smart agriculture case study. Section \ref{sec4} discusses the results and compares the proposed framework with existing approaches. Finally, Section \ref{sec5} concludes the paper and outlines future research directions.

\section{Background and Related Works}
\label{sec2}
This section overviews background concepts and highlights notable studies in threat analysis across other information security domains.
%%%%%%%%%%%%%%%%%%%%%%%%%%%%%%%%%%%%

\subsection{Cybersecurity Assessment of IoT Systems}
\label{sec2.1}

Cybersecurity assessment of IoT and other internet-connected systems involves the systematic evaluation of system architectures to identify potential vulnerabilities, attack vectors, and security risks \cite{ralston2007cyber}. Such assessment is essential for maintaining system integrity, preventing unauthorised access, and protecting sensitive data and operational functionality from cyber threats \cite{hussaini2022taxonomy}. Given the increasing integration of IoT technologies into safety-critical infrastructures, effective security evaluation has become a fundamental requirement for ensuring dependable system operation.

Security assessment can be conducted at multiple stages of the system lifecycle, including during conceptual design, system development, and post-deployment operation. Early-stage security analysis is particularly important because vulnerabilities introduced during the design phase may propagate throughout the system and become difficult to mitigate at later stages. Consequently, security analysis methods are increasingly applied during system design to evaluate potential risks and the effectiveness of existing security mechanisms \cite{abdulhamid2023overview,hussaini2021object}.

A comprehensive cybersecurity assessment typically involves identifying potential threats, analysing vulnerabilities, and evaluating the effectiveness of existing security controls. Based on this analysis, appropriate mitigation strategies can be developed, including the implementation of protective mechanisms such as encryption, authentication, access control policies, and network security measures \cite{aksu2017quantitative}. The effectiveness of such assessments depends largely on the ability to identify and prioritise risks according to their likelihood and potential impact \cite{ralston2007cyber}.

To support systematic risk identification and evaluation, structured modelling approaches are often employed to represent potential attack scenarios and system vulnerabilities. Among these approaches, threat modelling provides a formal mechanism for analysing how adversaries may exploit system weaknesses and how such attacks could propagate through interconnected components \cite{aksu2017quantitative}. Consequently, threat modelling techniques have become an important foundation for conducting rigorous cybersecurity analysis of IoT systems.

\subsection{Threat Modelling of IoT Systems}
\label{sec2.2}

Threat modelling is a fundamental component of cybersecurity assessment that systematically identifies, analyses, and evaluates potential threats and vulnerabilities within a system architecture \cite{mahak2021threat}. In IoT environments, threat modelling is particularly important due to the distributed nature of devices, heterogeneous communication protocols, and complex interactions between physical and digital components. These characteristics significantly expand the potential attack surface and introduce multiple pathways through which adversaries may compromise system functionality.

Threat modelling typically involves analysing the architecture, operational processes, communication protocols, and functional interactions of the system under consideration. By examining these elements, security analysts can identify potential weaknesses that attackers may exploit and evaluate how such vulnerabilities could affect system behaviour \cite{mahak2021threat}. The process often considers multiple attack scenarios, attacker capabilities, and the potential consequences of successful attacks on system integrity, availability, and confidentiality \cite{khan2017stride}.

The primary objective of threat modelling is to provide a structured understanding of cybersecurity risks in order to support the development of effective security policies and mitigation strategies. By identifying vulnerabilities and possible attack paths early in the system lifecycle, proactive measures can be implemented to reduce the likelihood and impact of cyber attacks.

Several threat modelling frameworks have been proposed to support systematic security analysis. One widely used approach is the STRIDE model, which categorises threats into six classes: Spoofing, Tampering, Repudiation, Information Disclosure, Denial of Service, and Elevation of Privilege \cite{khan2017stride}. Such frameworks provide structured guidance for identifying potential security threats and evaluating their potential impact. 

To further analyse how identified threats may propagate through a system and lead to compromise, formal modelling techniques are often employed. Among these techniques, Attack Tree Analysis provides a structured representation of attack paths and has been widely adopted for security risk modelling in complex systems.

\subsection{Model-Based Systems Engineering for Security Analysis}
\label{sec2.3.1}

Model-Based Systems Engineering (MBSE) has emerged as a systematic approach for managing the complexity of modern cyber-physical systems by representing system architectures, behaviours, and interactions using formal models rather than traditional document-based specifications. By providing structured representations of system components and their relationships, MBSE facilitates early-stage analysis, traceability, and verification throughout the system lifecycle.

In the context of IoT and other distributed cyber-physical systems, MBSE enables comprehensive modelling of heterogeneous components, communication interfaces, and operational dependencies. These capabilities are particularly valuable for security analysis, where vulnerabilities often arise from complex interactions between hardware, software, communication networks, and control processes. Integrating security analysis techniques within MBSE environments allows potential threats and vulnerabilities to be systematically identified and analysed at the architectural level.

Researchers, as reviewed in \cite{GRECHI2025112251}, have explored the use of MBSE to support safety and security co-analysis in complex systems, enabling designers to evaluate potential risks during early design stages and maintain traceability between system design artefacts and dependability assessments. By linking threat modelling techniques with system models, MBSE facilitates structured reasoning about how vulnerabilities in individual components may propagate through the system and lead to higher-level failures.

Within this context, ATA provides a suitable formalism for representing adversarial attack paths and analysing how system vulnerabilities may contribute to security breaches. When integrated with system architecture models, ATA enables systematic identification of attack scenarios and supports quantitative evaluation of cybersecurity risks in complex IoT environments.

\subsection{Attack Tree Analysis Model}
\label{sec2.3}

Attack Tree Analysis (ATA) is a graphical security modelling technique originally introduced by Bruce Schneier to systematically analyse potential attack scenarios against a system \cite{espedalen2007attack}. The method represents how an adversary may achieve a malicious objective by decomposing the attack goal into smaller sub-goals and basic attack steps \cite{45,mauw2006foundations,saini2008threat,ten2007vulnerability}. By structuring potential attack paths in a hierarchical form, ATA enables the identification of system vulnerabilities and the analysis of possible attack strategies that could lead to system compromise \cite{buldas2006rational}.

In an attack tree, the root node represents the attacker’s primary objective, commonly referred to as the TE. The TE typically denotes a successful system compromise or security-induced failure. This objective is decomposed into intermediate events (IEs), which represent attack goals or sub-goals, and Basic Attack Steps (BASs), which correspond to atomic adversarial actions required to achieve these goals \cite{roy2012attack,wen2021risk}. BASs represent the lowest level of the hierarchy and are often associated with specific vulnerabilities or exploit actions within the system.

\begin{figure}[t]
\centering
\includegraphics[scale=0.6]{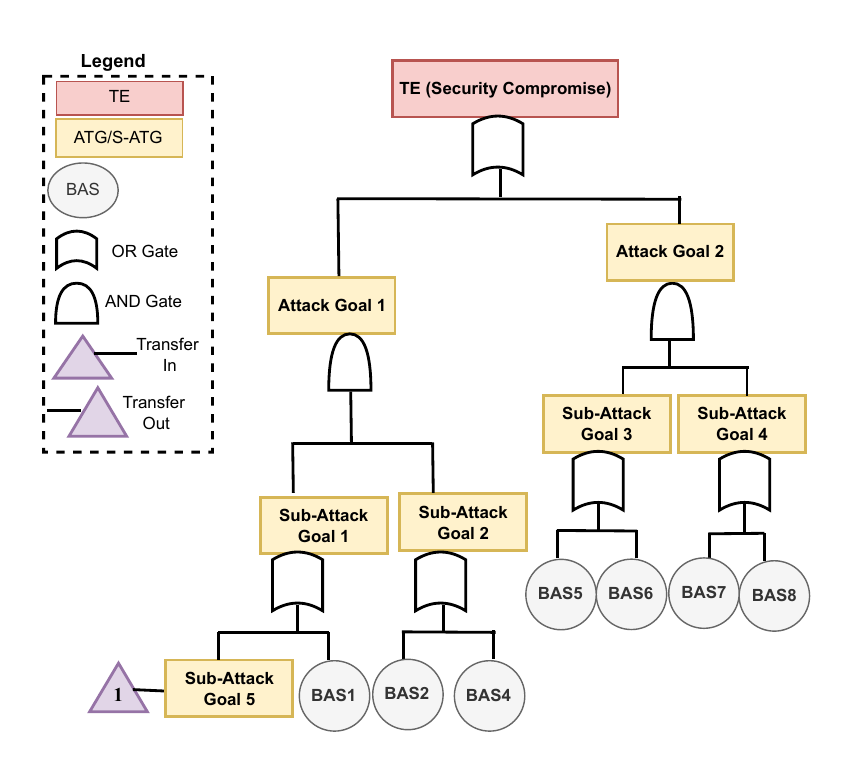}
\caption{An example attack tree}
\label{ATD}
\end{figure}

Logical relationships between nodes are represented using gate symbols, typically \textit{AND} and \textit{OR} gates. An \textit{AND} gate indicates that all child events must occur for the parent event to be activated, whereas an \textit{OR} gate indicates that the occurrence of any child event is sufficient to activate the parent node. These logical structures capture the dependencies among attack steps and enable systematic reasoning about how different attack paths may lead to system compromise. Figure~\ref{ATD} illustrates a representative attack tree structure, while Table~\ref{attack tree description} summarises the meaning of the nodes and gate symbols used in the model.

Quantitative ATA enables estimation of the likelihood of system compromise by assigning attribute values to BAS nodes \cite{55,45,mauw2006foundations,Reza2025}. These values are propagated through the tree using the logical relationships defined by the gates to compute the probability of the TE. Through this process, ATA supports the identification of critical attack paths and provides insights into the vulnerabilities that most significantly contribute to system-level security risk.

Due to its intuitive representation and analytical capabilities, ATA has been widely applied in information security, industrial control systems, and critical infrastructure protection \cite{55,45,mauw2006foundations,saini2008threat,ten2007vulnerability}. However, in many practical applications the probabilistic values assigned to BAS nodes rely heavily on expert judgement or qualitative scoring schemes, which may introduce subjectivity and uncertainty in the resulting risk assessment.

\begin{table}[!thpb]
\caption{Description of events and gates in an attack tree model}
\label{attack tree description}
\centering
\begin{tabular}{*{15}{l}}
\hline
  \textbf{Events/Gates} & \textbf{Description} 
\tabularnewline
    \hline
TE & Security-induced failure of the system\\
IEs & Attack and sub-attack goals  \\
BASs & \makecell[l]{The atomic attack steps are represented \\as vulnerability nodes}\\
\hline
AND Gate & \makecell[l]{All inputs must happen for a successful \\attack} \\
OR Gate & \makecell[l]{At least one input will cause a successful \\attack} \\
\hline 
\end{tabular}
\end{table}

%The ATA model calculates the likelihood of system-level failure caused by external malicious activities. An attribute value is assigned to each BAS to obtain this information \cite{55,45,mauw2006foundations,Reza2025}. These values are combined using the boolean gates to quantify the overall system-level security risk probability. The security risk probabilities of a system are calculated by considering the probability of each event in the hierarchy of the tree from bottom to up. 

%Using the ATA model, the risk associated with each attack path can be assessed, the most critical ones prioritised, and addressed proactively. This helps mitigate the risks quantitatively and protect against potential threats. The model is an essential element of a comprehensive security framework and is widely used in various fields, including information security, industrial control systems, and critical infrastructure protection \cite{55,45,mauw2006foundations,saini2008threat,ten2007vulnerability}.

%%%%%%%%%%%%%%%%%%%%%%%%%%%%%%%%%%%%%%%%%%%%%%%%%%%%%%%%%%%%%%%%%%%%%%%%%%%%%%%%%%%%%%
\subsection{Exploit Prediction Scoring System}
\label{sec2.4}

The Exploit Prediction Scoring System (EPSS) is a data-driven framework introduced by the Forum of Incident Response and Security Teams (FIRST) to estimate the likelihood that a disclosed software vulnerability will be exploited in the wild \cite{jacobs2023enhancing}. Unlike traditional vulnerability scoring approaches that primarily assess severity, EPSS focuses on predicting real-world exploitation probability, thereby supporting more informed vulnerability prioritisation and remediation decisions.

EPSS employs statistical modelling and machine learning techniques to analyse historical exploitation data and vulnerability characteristics in order to estimate the probability that a specific vulnerability will be exploited within a given time horizon \cite{jacobs2023enhancing}. The framework integrates information from multiple sources, including publicly disclosed vulnerability records, exploit databases, security advisories, and threat intelligence reports \cite{jacobs2023enhancing}. By analysing patterns in past exploitation behaviour, EPSS generates probabilistic scores ranging from 0 to 1 that indicate the likelihood of real-world exploitation.

Although EPSS was inspired by the Common Vulnerability Scoring System (CVSS), the two frameworks serve different purposes. CVSS primarily evaluates the technical severity of vulnerabilities, whereas EPSS estimates their probability of exploitation. Consequently, EPSS provides complementary information that can support more effective vulnerability prioritisation and risk assessment \cite{jacobs2021exploit}. 

Because EPSS provides regularly updated probabilistic exploitability estimates derived from empirical data, it offers a valuable resource for quantitative cybersecurity modelling. In particular, EPSS probabilities can be integrated into security analysis frameworks to assign objective likelihood values to vulnerabilities, thereby reducing reliance on subjective expert judgement. This capability makes EPSS particularly suitable for supporting data-driven cybersecurity assessment of complex IoT-based systems.

\subsection{Related Works}
\label{sec2.5}

The rapid adoption of IoT and UAV technologies has enabled significant advancements in applications such as precision agriculture, environmental monitoring, and industrial automation. In smart agriculture in particular, IoT-based sensing and UAV-assisted monitoring have been widely employed to support crop management, irrigation control, and disease detection \cite{maslekar2020application,kumar2020smart,dahane2020iot,gutierrez2013automated}. Despite these benefits, the increasing reliance on interconnected IoT infrastructures introduces significant cybersecurity and dependability challenges. Failures or malicious exploitation of sensor-based IoT devices may lead to severe operational disruptions, environmental damage, and reduced agricultural productivity. Such risks highlight the importance of incorporating robust security and dependability assessment mechanisms into IoT-based systems \cite{abdulhamid2022dependability,abdulhamid2023overview}.

The resource-constrained and distributed nature of IoT networks further complicates the implementation of effective security mechanisms. Limited computational capabilities, heterogeneous communication protocols, and large-scale device deployment create a broad attack surface that can be exploited by adversaries \cite{hammoudeh2019continuous,xing2020}. Consequently, numerous studies have investigated security challenges in IoT environments and proposed mechanisms for mitigating potential risks \cite{hammoudeh2019continuous,lawal2020security,kumar2016security,mosenia2016comprehensive}. 

Several works have explored quantitative dependability and security modelling techniques for IoT infrastructures. For example, Nguyen \textit{et al.} \cite{nguyen2021dependability} proposed a hierarchical Fault Tree Analysis (FTA) framework to evaluate reliability and cybersecurity risks in Internet-of-Medical Things (IoMT) infrastructures by modelling multiple failure modes, including cyber attacks on software subsystems. Similarly, Nguyen \textit{et al.} \cite{nguyen2020hierarchical} developed a quantitative modelling framework combining Reliability Block Diagrams (RBD), Fault Trees (FT), and Continuous-Time Markov Chains (CTMC) to analyse the availability and security of IoT infrastructures.

Within the cybersecurity domain, Attack Tree Analysis has been widely adopted for modelling adversarial behaviour and analysing attack paths. Foundational studies by Asif \textit{et al.} \cite{45}, Mauw \textit{et al.} \cite{mauw2006foundations}, Ten \textit{et al.} \cite{ten2007vulnerability}, Saini \textit{et al.} \cite{saini2008threat}, and Roy \textit{et al.} \cite{roy2012attack} introduced various extensions of the ATA framework, including quantitative attack trees, Attack Countermeasure Trees (ACTs), and Quantitative Attack-Defence Trees (QADT). These approaches enable structured modelling of adversarial strategies and provide mechanisms for analysing potential attack paths in cyber-physical and industrial control systems.

%Several studies have further applied ATA-based approaches to evaluate system security. 
Asif \textit{et al.} \cite{45} developed an attack tree model incorporating attacker motivation, computational resources, and attack attributes to estimate attack risk probabilities. Mauw \textit{et al.} \cite{mauw2006foundations} established theoretical foundations for quantitative attack tree modelling and proposed methods for analysing attack attributes within the tree structure. Ten \textit{et al.} \cite{ten2007vulnerability} applied ATA models to evaluate vulnerabilities in Supervisory Control and Data Acquisition (SCADA) systems, demonstrating their usefulness in identifying potential break-in points and security weaknesses. Saini \textit{et al.} \cite{saini2008threat} proposed a versatile ATA-based threat modelling framework for analysing cyber threats and strengthening security mechanisms. Roy \textit{et al.} \cite{roy2012attack} introduced ACTs to represent both attack and defence mechanisms within the modelling framework. Other works, such as Kumar \textit{et al.} \cite{kumar2015quantitative} and Foster \textit{et al.} \cite{forster2010integration}, utilised ATA models to evaluate attack paths and estimate the resources required for successful attacks using expert-derived parameters. More recently, Abdulhamid \textit{et al.} \cite{abdulhamid2025quantitative} incorporated fuzzy set theory and expert judgement within ATA models to estimate attack probabilities and system failure risks.

Although these approaches provide valuable insights into cyber threat modelling, many rely on subjective estimations or expert judgement when assigning probability values to attack events. Such subjective quantification can introduce uncertainty and limit reproducibility in cybersecurity risk assessment.

Alternative approaches have attempted to incorporate vulnerability severity metrics into probabilistic models. For instance, Flores \textit{et al.} \cite{flores2022smart} proposed a Bayesian Network model that utilises vulnerability severity scores from the CVSS to analyse the security of IoT-based smart homes. While the approach considers attack scenarios such as Denial-of-Service and man-in-the-middle attacks, reliance on severity-based metrics alone may not accurately reflect the real-world likelihood of vulnerability exploitation.

Recent developments in vulnerability intelligence, particularly the EPSS, provide probabilistic estimates of real-world exploitability based on empirical evidence \cite{jacobs2021exploit}. These exploitability scores offer an opportunity to support objective and data-driven cybersecurity modelling by assigning evidence-based probabilities to vulnerabilities.

Motivated by these limitations, this study integrates EPSS exploitability probabilities within an Attack Tree Analysis framework to enable data-driven quantification of cybersecurity risks in IoT systems. In the proposed approach, identified vulnerabilities are mapped to BASs within the attack tree structure, and EPSS-derived probabilities are used to quantify the likelihood of successful exploitation. This enables systematic estimation of system-level compromise probability while reducing reliance on subjective probability assignment.

\section{Methodology: EPSS-Integrated Quantitative Attack Analysis Framework}
\label{sec3}

\subsection{Overview of the Framework}

This study proposes a structured methodology for quantitative cybersecurity risk assessment that integrates MBSE, ATA, empirical vulnerability exploitability data, and Bayesian inference within a unified analytical framework. The approach transforms an architectural system model into analysable security artefacts, enabling systematic estimation of system-level compromise probabilities. In particular, the framework derives attack tree models from system architecture representations and subsequently transforms them into a BN, supporting both deterministic evaluation and probabilistic inference-based cybersecurity analysis.

The MBSE-based attack tree generation builds upon our previous work \cite{abdulhamid2025b}, which focused on modelling system reliability under random component and communication failures. In that work, failure behaviour was represented using exponentially distributed component failure probabilities and cyber attacks were not considered. The current study extends this modelling approach by explicitly incorporating cybersecurity threats into the analysis. Unlike random hardware failures, cyber attacks cannot be realistically characterised using exponential failure distributions. Therefore, this work integrates empirically derived exploitability probabilities obtained from the EPSS, enabling data-driven quantification of vulnerability exploitation likelihood.

\begin{figure*}[!thpb]
\centering
\includegraphics[scale=0.9]{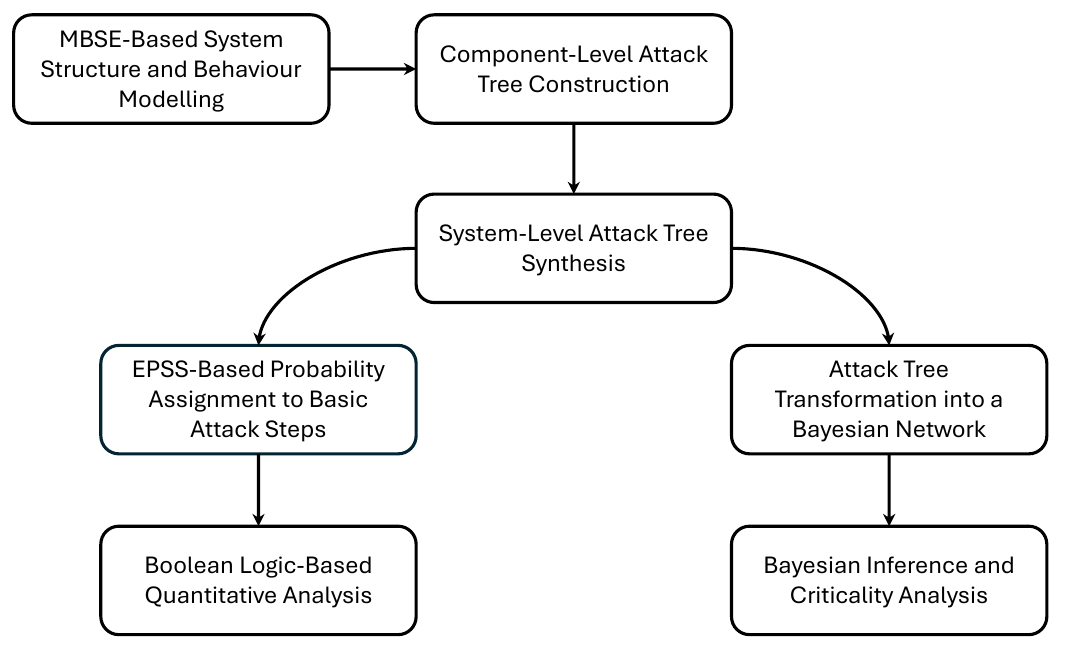}
\caption{ Steps of the Proposed Framework}
\label{propfr}
\end{figure*}

The overall methodological pipeline, illustrated in Figure~\ref{propfr}, consists of six main stages: (i) system behaviour modelling within an MBSE environment; (ii) transformation of behavioural models into component-level attack trees; (iii) synthesis of a system-level attack tree using architectural connectivity; (iv) assignment of empirical exploitability probabilities derived from EPSS; (v) quantitative propagation of probabilities through Boolean logic gates to estimate system compromise likelihood; and (vi) transformation of the attack tree into a Bayesian Network to enable probabilistic inference and vulnerability criticality analysis. The formal computational procedure corresponding to this workflow is summarised in Algorithm~\ref{algorithm}.

By combining architectural modelling with empirical vulnerability intelligence, the proposed framework reduces the reliance on subjective probability assignment commonly found in traditional attack tree quantification approaches. This integration enables a more objective estimation of system-level breach probabilities and supports the identification and prioritisation of critical vulnerabilities for cybersecurity mitigation.

\subsubsection{Methodological Assumptions}

The proposed framework relies on several modelling assumptions to enable quantitative cybersecurity analysis:

\begin{itemize}
\item Vulnerabilities identified in the system architecture are represented as BASs in the attack tree structure.

\item EPSS exploitability scores are treated as probabilistic estimates of successful vulnerability exploitation and are used as likelihood values for corresponding BAS nodes.

\item Attack events represented in the attack tree are assumed to be conditionally independent unless explicitly linked through logical gate structures.

\item Successful exploitation of a vulnerability is assumed to result in the compromise or failure of the associated system component within the attack model.

%\item The system architecture represented in the MBSE model accurately captures component interactions and communication pathways that may enable attack propagation.
\end{itemize}

These assumptions enable the systematic integration of architectural modelling, vulnerability intelligence, and probabilistic reasoning within the proposed cybersecurity quantification framework. The individual stages of the framework are detailed in the following subsections.

\begin{algorithm}
\caption{EPSS-Integrated MBSE Attack Quantification}
\begin{algorithmic}[1]
\Require SysML model $M$, EPSS dataset
\Ensure $Pr(TE)$, vulnerability ranking
\State Extract components from BDD
\State Extract connectivity from IBD
\State Extract failure states from SMD
\For{each component $C_i$}
    \State Generate component attack tree
\EndFor
\State Integrate into system attack tree (SAT)
\State Assign EPSS probabilities to BASs
\State Compute $Pr(TE)$ via Boolean propagation
\State Transform SAT to Bayesian Network
\State Perform BN Inference and Compute Birnbaum Importance for each BAS
\State Return $Pr(TE)$ and vulnerability ranking
\end{algorithmic}
\label{algorithm}
\end{algorithm}

\subsection{Behaviour Modelling of the IoT System}
The behaviour modelling stage uses SysML to formally represent the IoT system architecture, extending the approach of \cite{abdulhamid2025b} by incorporating cyber threat considerations. Both structural and behavioural aspects are captured to support subsequent attack tree generation.

\subsubsection{Static Structure Modelling}
The static configuration of the system is represented using SysML Block Definition Diagrams (BDDs), which represent the hierarchical decomposition of the system into hardware and software components. The BDD defines associations, generalisations, dependencies, aggregations, and inheritance relationships, thereby establishing a formal architectural taxonomy. While \cite{abdulhamid2025b} focused solely on reliability-relevant components, the current framework also identifies components susceptible to cyber compromise, ensuring that potential attack surfaces are represented at the architectural level.

Formally, the IoT system is represented as:
\begin{equation*}
    S=\{C_1, C_2, \cdots, C_n\}
\end{equation*}

where each component $C_i$ may contain multiple vulnerability-relevant states or attack vectors as:

\begin{equation*}
    C_i=\{v_{i1}, v_{i2}, \cdots, v_{im}\}
\end{equation*}

where $v_{ij}$ represents exploitable vulnerabilities or attack-relevant failure modes.

This architectural abstraction provides the structural basis for modular attack tree generation.

\subsubsection{Internal Structure Modelling}

Internal structure modelling is conducted using Internal Block Diagram (IBD). The IBD specifies component instances, ports, connectors, and data flows, thereby illustrating how system elements interact and exchange information. Modelling internal interactions is critical for IoT systems, where communication pathways and interdependencies significantly influence attack propagation. The connectivity extracted from the IBD later determines the logical gate configuration when synthesising the system-level attack tree. The inclusion of potential attack paths differentiates this work from \cite{abdulhamid2025b}, providing the basis for attack tree synthesis at both component and system levels. 

\subsubsection{Nominal and Failure Behaviour Modelling}

Nominal and failure behaviours are modelled using State Machine Diagrams (SMDs). Each component’s operational logic is represented as a finite state transition system, where nominal states correspond to functional behaviour and failure or compromised states are derived using a lightweight extension of the SysML meta-model through DAM stereotypes (DaStep) that support the explicit representation of attack-induced failure behaviours. 

 These annotated failure states represent potential BASs in the generated attack tree. As this article focuses solely on the cyber attack-induced system failures, this ensures that the resulting attack tree encompasses security considerations. The transition logic encoded in the SMD determines whether exploitation paths are conjunctive or disjunctive, thereby influencing subsequent logical gate selection.

\subsection{Transformation of the MBSE Source Model into Attack Tree Artefacts}

The transformation process converts the annotated SysML source model into formal attack tree representations suitable for quantitative analysis. A pattern-based transformation approach is adopted to systematically map behavioural elements to attack tree constructs.

This transformation is executed in two stages: first, component-level attack trees are generated; second, these are integrated into a system-level attack tree using architectural connectivity information.

\subsubsection{Component-Level Attack Tree Generation}
Each component’s annotated SMD is transformed into a Component Attack Tree (CAT), following the procedure in \cite{abdulhamid2025b} but extended to include cyber attacks. Failure or compromised states are mapped to BASs, and transitions among these states determine the logical relationships between nodes.

Boolean logic gates (AND/OR) define how basic events propagate to component-level compromises. If multiple vulnerabilities independently lead to compromise, they are connected through an \textit{OR} gate. If multiple conditions must be satisfied jointly, they are connected via an \textit{AND} gate. This transformation ensures traceability from behavioural modelling artefacts to quantitative attack logic.

While \cite{abdulhamid2025b} assumed exponential failure probabilities for the Basic Events, in this framework, however, each BAS corresponding to a cyber attack is mapped to a Common Vulnerabilities and Exposures CVE entry, and its probability is assigned using the EPSS. Thus,
\begin{equation}
    Pr(BAS_i)=EPSS(CVE_i)
    \label{EPSS}
\end{equation}
%BASs corresponding to cyber attacks are quantified using EPSS data, capturing real-world exploit likelihoods. 

\subsubsection{ System-Level Attack Tree Synthesis}

The overall system attack tree is constructed by integrating component-level attack trees using connectivity information derived from the IBD. Architectural configurations determine gate semantics. In series configurations, compromise of any component may propagate upward, and therefore OR gates are employed. In parallel or redundant configurations, AND gates represent the requirement for simultaneous compromise. By integrating attack paths, the SAT provides a holistic view of system vulnerability. Compared to \cite{abdulhamid2025b}, this stage introduces cyber threats as additional top-level contributors, enabling security risk assessment. The TE is defined as system-level cybersecurity breach. The resulting hierarchical tree captures propagation of exploitation from leaf-level BAS nodes to the TE.

\subsection{Quantitative Probability Assignment and Propagation via Boolean Logic}

For quantitative evaluation, cyber attack-induced component failures are treated with EPSS-based exploit probabilities. EPSS data provides empirically grounded probabilities for cyber attack success, reflecting real-world threat likelihoods rather than purely theoretical distributions. Boolean propagation is applied across the attack tree to compute system-level probabilities of compromise, supporting deterministic evaluation of potential breaches.

The TE probability is computed using recursive bottom-up propagation. For an OR gate with input events $BAS_i$:

\begin{equation}
    Pr_{OR}=1-\prod_{i=1}^{N}{\big( 1-Pr(BAS_i)\big)}
\label{OR}
\end{equation}

For an AND gate:

\begin{equation}
    Pr_{AND}=\prod_{i=1}^{N}{ Pr(BAS_i)}
    \label{AND}
\end{equation}

This deterministic gate-by-gate computation yields the system-level breach probability under the independence assumption.

\subsection{Bayesian Transformation and Criticality Analysis}

To enhance analytical flexibility and enable probabilistic reasoning, the attack tree is transformed into a Bayesian Network. During graphical mapping, BASs are mapped to root nodes, intermediate events to intermediate nodes, and the TE to a leaf node. During numerical mapping, EPSS-based probabilities define priors for root nodes, and conditional probability tables are derived from Boolean gate semantics.

The BN representation enables posterior inference, predictive risk assessment, and criticality analysis using measures such as the Birnbaum Importance Measure (BIM), identifying which components or attacks contribute most significantly to system risk. The BIM is computed by evaluating the change in TE probability when a BAS is forced to failure versus non-failure. This ranking supports vulnerability prioritisation and design optimisation. This approach extends \cite{abdulhamid2025b} by integrating data-driven cyber attack probabilities and enabling probabilistic risk reasoning alongside traditional reliability assessment.

 \section{Case Study: Smart Agriculture IoT System as Demonstrative Validation}
 \label{casestudy}

The smart agriculture IoT system is employed as a demonstrative case study to validate the applicability and scalability of the proposed EPSS-integrated model-based quantitative attack analysis framework. The objective is not to provide a domain-specific security evaluation of agricultural systems, but rather to demonstrate how an IoT architecture can be systematically transformed from a SysML source model into quantitative cybersecurity analysis artefacts through a structured MBSE-to-ATA-to-BN pipeline.

The case study therefore serves as a validation instance of the general framework introduced in Section \ref{sec3}, illustrating how architectural abstraction, behavioural modelling, vulnerability annotation, and empirical exploitability data can be integrated into a unified analytical workflow.

%The proposed model for cyber-attack-induced failure probability analysis of IoT-based systems involves five detailed steps as shown in Figure \ref{propfr}. The framework involves system description, creating an attack tree model, identification of vulnerabilities and their EPSS probabilities, identification of attacks and sub-attack goals and security quantification of the system.

\subsection{Case Study System Description}
\label{sec3.1}

The IoT-enabled Smart Irrigation System (SIS) is used as a running case study to demonstrate the applicability of the proposed quantitative cybersecurity modelling framework. The SIS architecture was previously introduced in our earlier studies \cite{Mokhlesur2023,abdulhamid2024enhsaf}, where the emphasis was placed on qualitative safety assessment and architectural dependability analysis. In contrast, the present work extends that line of research by focusing explicitly on quantitative cybersecurity modelling and security-induced system failure analysis.

As outlined in Section \ref{sec1}, the overall dependability of the SIS encompasses safety, reliability, security, and maintainability. Existing research in the broader dependability domain has predominantly concentrated on safety and reliability modelling \cite{xing2020,steiner2015qualitative,LGEN21,kabir2015reliability,abdulhamid2023overview,abdulhamid2023adaptation,rahman2023qualitative,abdulhamid2024reliability}, leaving a clear methodological gap in the rigorous quantification of cybersecurity risks, particularly for IoT-based smart irrigation infrastructures. This study addresses that gap by applying the proposed MBSE-driven attack tree and Bayesian inference framework to the SIS, enabling empirical exploitability-based breach estimation.

At a high level, the SIS leverages IoT technologies to enable remote supervision and control of agricultural irrigation processes. The system allows farmers to monitor environmental conditions in real time and make informed irrigation decisions, ensuring that crops receive adequate water while optimising resource utilisation. By integrating distributed sensing, wireless communication, and edge/cloud processing, the SIS aims to maximise crop yield, conserve water, and reduce operational costs associated with manual monitoring.

\begin{figure}[!thpb]
\centering
\includegraphics[scale=0.30]{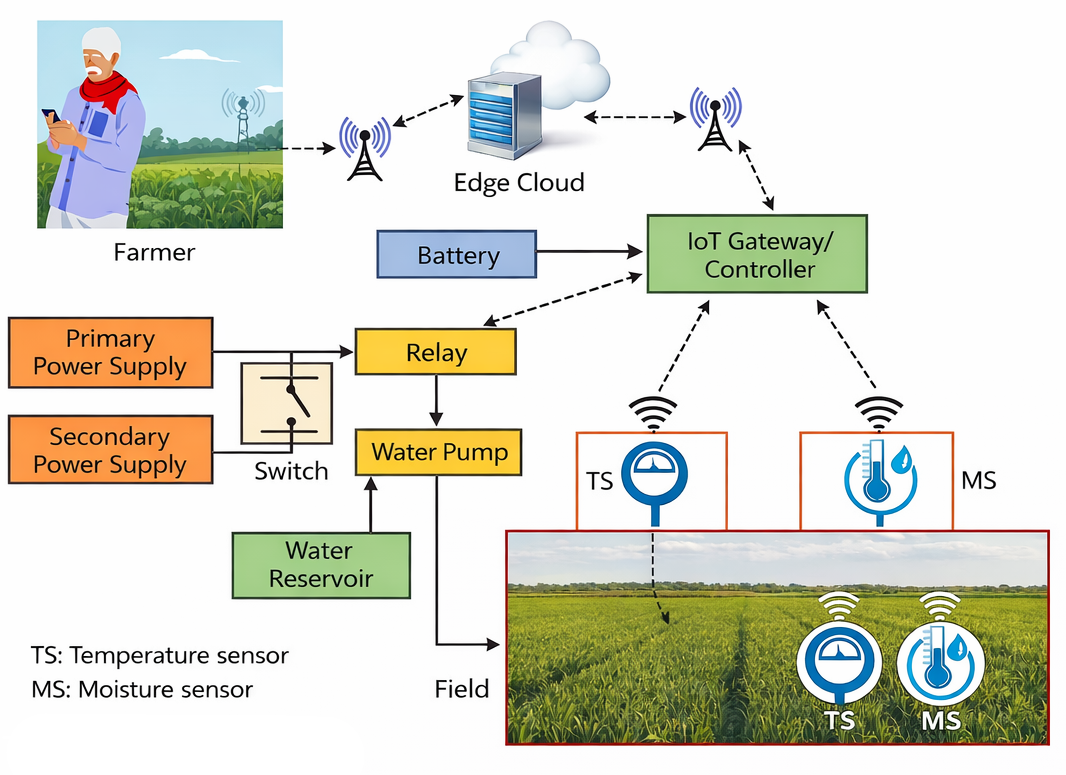}
\caption{ IoT-based smart irrigation environment (adapted from \cite{abdulhamid2024enhsaf})}
\label{fig1}
\end{figure}

Operationally, the system functions as illustrated in Figure \ref{fig1}. Environmental data are collected from the field using two primary sensors: a temperature sensor and a soil moisture sensor. These sensors continuously measure field conditions and transmit the acquired data wirelessly to an IoT gateway (controller). The gateway performs signal conditioning and converts the received analogue signals into digital representations suitable for computational processing. The processed data are then forwarded to an edge cloud server for decision support.

The edge cloud server incorporates an expert-system-based decision engine that analyses incoming environmental data against predefined irrigation rules. Based on soil moisture thresholds and temperature conditions, the system determines whether irrigation is required and, if so, specifies the duration of water delivery. Once a decision is generated, the information is communicated to the farmer via wireless communication channels.

The farmer retains supervisory control and can choose to activate or deactivate the irrigation pump. If activation is selected, a control signal is transmitted back through the communication infrastructure to the edge server, which instructs the IoT gateway to energise the relay controlling the sprinkler system. The sprinkler system is thus actuated to irrigate the field for the specified duration. For the purpose of this analysis, the power supply subsystem is assumed to be uninterrupted, supported by both primary and backup energy sources.

From a cybersecurity perspective, the SIS presents multiple potential attack surfaces, including wireless sensor communications, gateway firmware, edge cloud processing, and remote control signalling. These interconnected components make the system an appropriate and representative case study for evaluating the proposed MBSE-based attack tree synthesis and EPSS-driven quantitative breach estimation framework.

\subsection{MBSE-Based Transformation for Cybersecurity Analysis}

The architectural representation of the SIS remains unchanged from our previous work \cite{abdulhamid2024enhsaf}, where the system was modelled for safety and reliability assessment. Since the BDD and IBD capture structural composition and inter-component data flow independent of the analysis objective, these artefacts are reused in the present study.

\begin{figure*}[!thpb]
\centering
\includegraphics[scale=0.35]{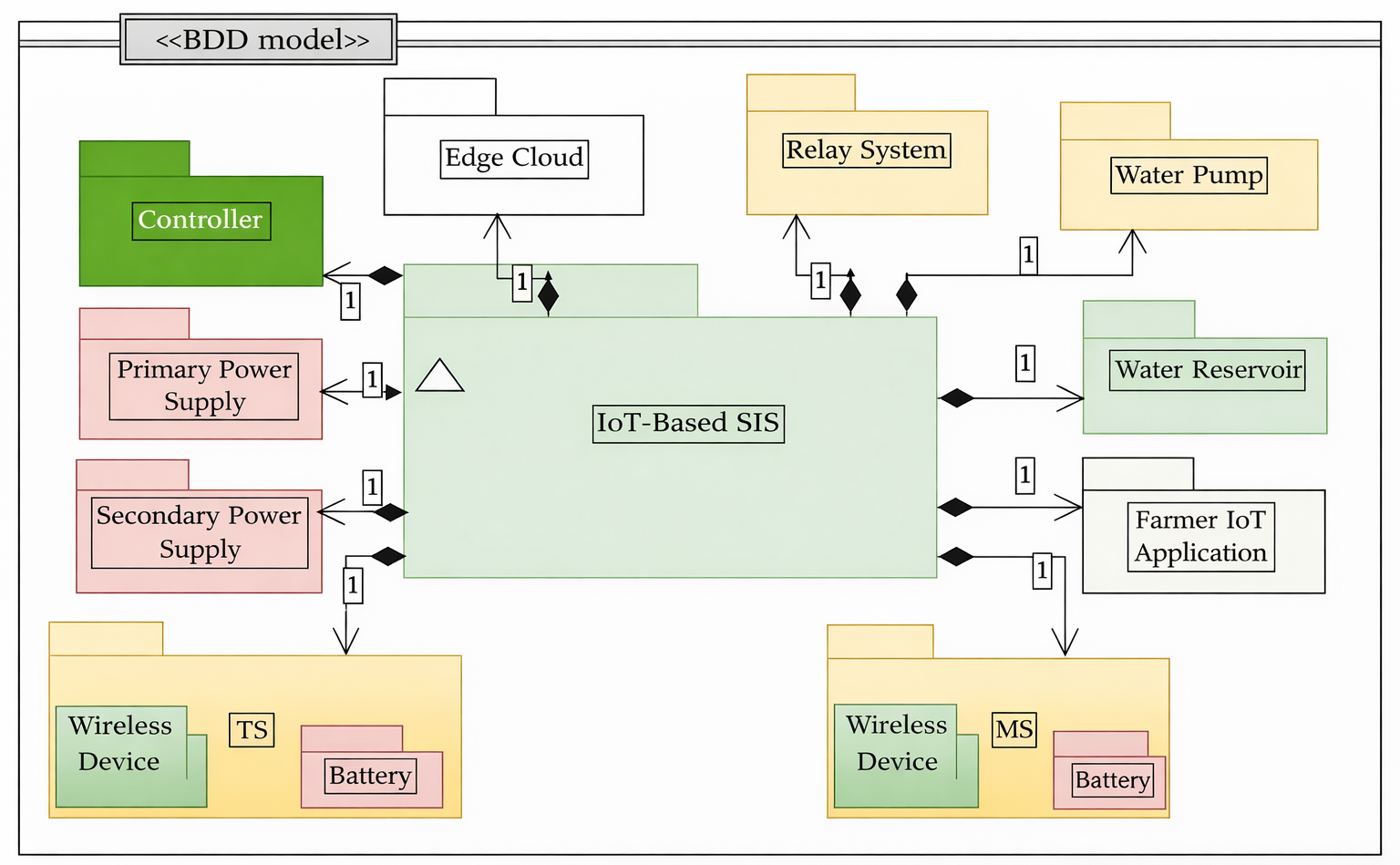}
\caption{ BDD of the SIS as seen in \cite{abdulhamid2024enhsaf}}
\label{fig-BDD}
\end{figure*}

\begin{figure*}[!thpb]
\centering
\includegraphics[scale=0.75]{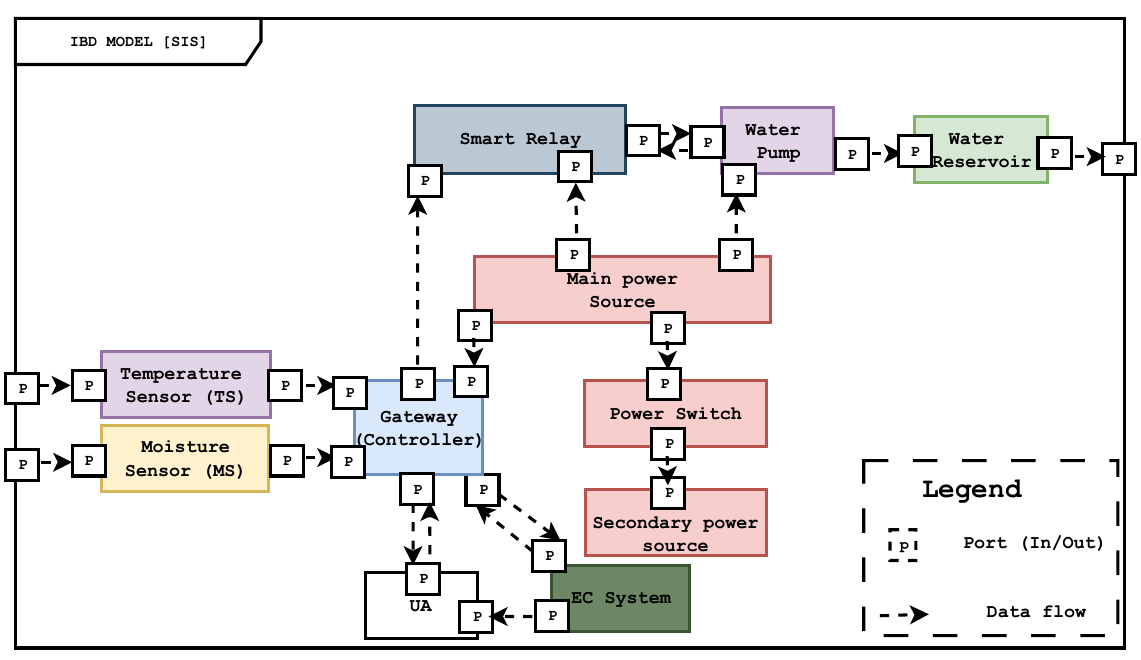}
\caption{ IBD of the SIS as seen in \cite{abdulhamid2024enhsaf}}
\label{fig-IBD}
\end{figure*}

The BDD (Figure \ref{fig-BDD}) formally specifies the hierarchical decomposition of the SIS into its functional components and composite blocks, establishing composition relationships and structural dependencies. The IBD (Figure \ref{fig-IBD}) captures the internal configuration of the system, representing ports, connectors, and item flows that define communication paths between sensors, gateway, edge server, control unit, and actuation mechanisms. As attack propagation is inherently dependent on architectural connectivity, the reuse of these diagrams ensures structural consistency while enabling cybersecurity-oriented modelling.

Unlike \cite{abdulhamid2024enhsaf}, however, the analytical objective in this paper is not safety-induced system failure but security-induced compromise. Therefore, while the architectural foundation is preserved, the behavioural annotation and transformation logic are extended to incorporate cyber attack scenarios.

In \cite{abdulhamid2024enhsaf}, failure annotation was performed using SysML SMDs extended with the DAM profile to represent random component failures. Failure states such as “no sensor output” or “power source failure” were triggered by reliability-related events, and transitions were annotated accordingly.

In the present study, this modelling approach is extended to represent cyber attack-induced failure behaviour. Instead of random failure triggers, state transitions now include cyber compromise events derived from identified vulnerabilities. The DAM stereotype mechanism is retained for consistency; however, failure transitions are reinterpreted to capture malicious exploit activation rather than stochastic hardware failure.

\begin{figure*}[!thpb]
\centering
\includegraphics{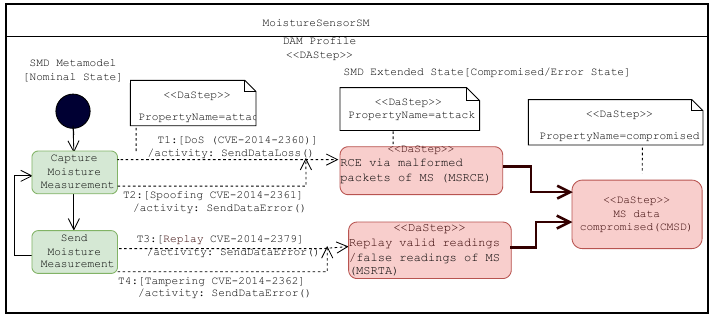}
\caption{Attack-induced failure behaviour annotation with an SMD}
\label{fig-SMD}
\end{figure*}

To illustrate this extension, a security-annotated SMD is provided in Figure \ref{fig-SMD} for the moisture sensor component. The nominal operational state represents correct sensing and transmission of soil moisture data. Additional states are introduced to capture attack-induced deviations. Transitions to these states are triggered by exploit events associated with identified vulnerabilities of the sensor firmware or communication interface. Each exploit event is later mapped to an EPSS probability value for quantitative analysis.

The same annotation procedure can be systematically applied to other components in the SIS, including the gateway, edge server, and communication links. For brevity, only one representative component is illustrated, as the transformation rules remain uniform across the system.

\subsection{Attack Modelling and Vulnerability Quantification}

\subsubsection{Creation of the Attack Tree Model from System Architecture}

Following the security-oriented behavioural annotation of the SIS components, the MBSE artefacts are transformed into an Attack Tree (AT) representation. Unlike the fault tree generation in \cite{abdulhamid2024enhsaf}, where random component and communication failures were modelled, the present work considers deliberate cyber attacks as causal mechanisms leading to system compromise. The Top Event (TE) of the attack tree is defined as: \textit{``Failure of SIS due to deliberate cyber attacks.''}

The attack tree adopts a hierarchical structure, modelling how adversarial actions propagate through system components to realise this undesired event. The structure of the tree is derived from the architectural dependencies captured in the BDD and IBD models and from the attack-induced states introduced in the security-annotated SMDs.

\begin{figure*}[!thpb]
\centering
\includegraphics[scale=0.5]{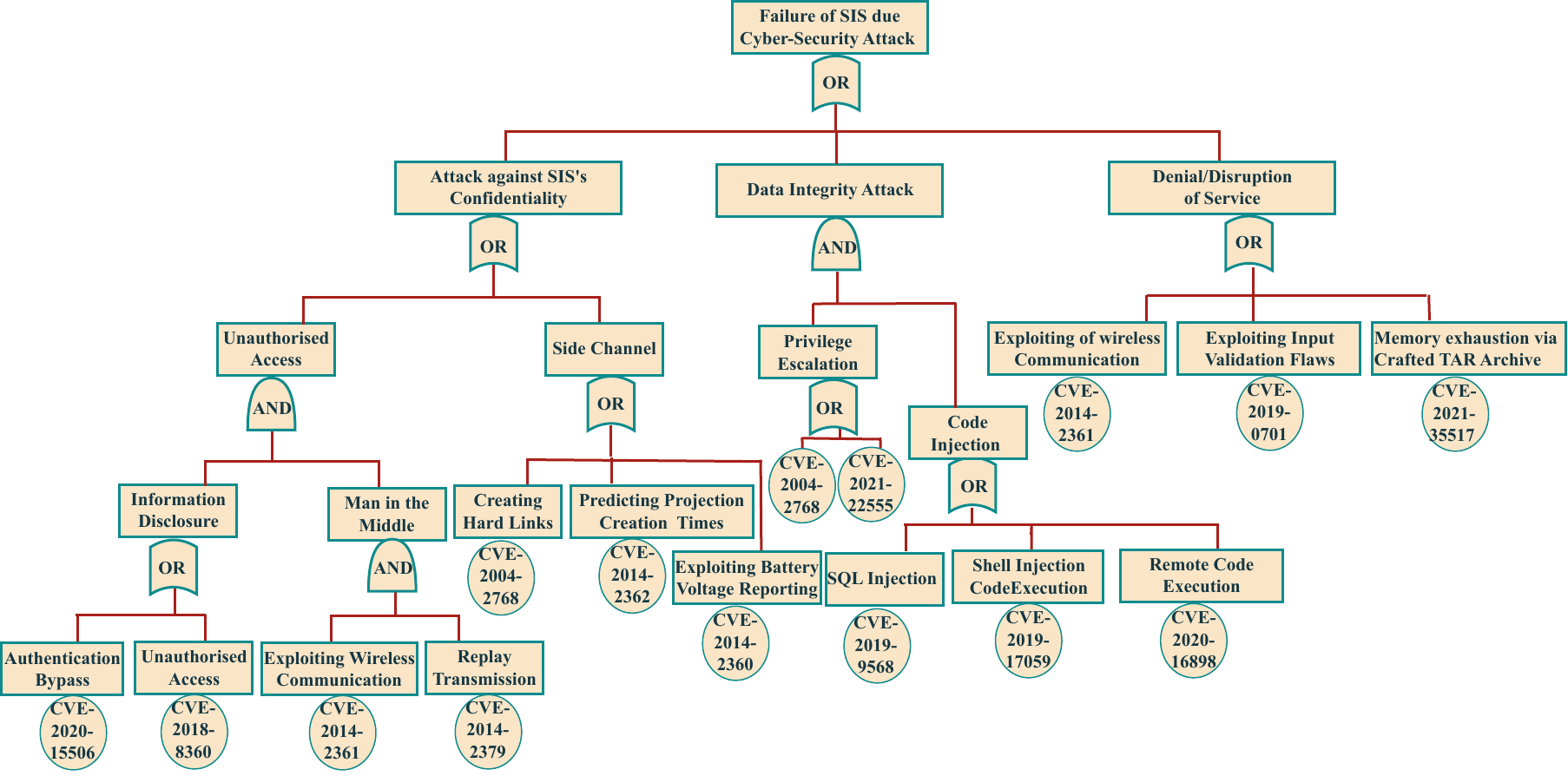}
\caption{ Attack tree of the IoT-based smart irrigation environment}
\label{description:attacktree}
\end{figure*}

The resulting minimised System-Level Attack Tree (SAT), shown in Figure \ref{description:attacktree}, captures attack paths from vulnerability exploitation at the component level to system-wide irrigation failure. For brevity, intermediate component-level attack trees are not presented; however, their construction follows the deterministic SMD-to-attack-tree transformation process described earlier. The figure therefore represents a minimised system-level abstraction suitable for quantitative evaluation.

%Attack tree modelling effectively captures the cyber attack-induced failure behaviour of a system using a top-down approach representing security events and logical gates, as outlined in Section \ref{sec2.3}. The analysis focuses specifically on IoT system failures due to security breaches. Using ATA model, the top undesired event identified as the TE in the ATA model is described as . This threat event and its potential occurrences are graphically depicted in the ATA model in Figure \ref{description:attacktree}.

\subsubsection{Structuring of Attack Goals}
The intermediate nodes of the attack tree are organised according to the classical Confidentiality, Integrity, and Availability (CIA) security objectives. These represent sub-attack goals whose successful realisation may lead to the TE.
\begin{itemize}
    \item \textbf{Confidentiality compromise} includes unauthorised access, authentication bypass, and information disclosure.
    
    \item \textbf{Integrity compromise} includes code injection, command execution, privilege escalation, and manipulation of sensor or control data.
    
    \item \textbf{Availability compromise} includes denial-of-service (DoS) conditions and communication disruption.
\end{itemize}

This structuring ensures that the attack tree reflects not only architectural dependencies but also security semantics aligned with recognised threat modelling principles.

\subsubsection{Identification of Vulnerabilities as Basic Attack Steps (BAS)}

Each leaf node of the attack tree, referred to as a BAS, corresponds to a publicly disclosed vulnerability identified using the CVE system. These vulnerabilities are selected based on the architectural characteristics of the SIS components, including sensors, gateway, edge server, communication interfaces, and control subsystems.

Threat identification is performed using the STRIDE methodology \cite{khan2017stride}, enabling systematic mapping between attack types and component-level weaknesses. The analysis assumes that a malicious actor may exploit any identified vulnerability to compromise the system. By taking into account factors such as attack complexity, likelihood of discovery, potential impact, and current threat landscape, numerical insights of EPSS can be generated from the CVE system \cite{janiszewski2022creating,ali2011new,scarfone2009analysis,jacobs2021exploit}

Table \ref{tab:epss} summarises the selected vulnerabilities and their associated EPSS values obtained from the NVD database \cite{nvd}. Each CVE entry is mapped directly to a BAS node in the attack tree. The impact column categorises vulnerabilities according to their CIA effect.

Examples include:
\begin{itemize}
    \item \textbf{Authentication bypass and unauthorised access} (CVE-2020-15506, CVE-2018-8360), enabling confidentiality breaches.
    \item \textbf{Remote Code Execution} (CVE-2020-16898) and \textbf{Shell Injection} (CVE-2019-17059), enabling integrity compromise through arbitrary command execution.
    \item \textbf{SQL Injection} (CVE-2019-9568), enabling database manipulation.
    \item \textbf{Denial-of-Service vulnerabilities} (e.g., CVE-2019-0701, CVE-2021-35517), enabling availability disruption.
    \item \textbf{Man-in-the-Middle vulnerabilities} (CVE-2014-2379, CVE-2014-2361), enabling communication interception and manipulation.
\end{itemize}

By explicitly mapping CVEs to BAS nodes, the attack tree becomes directly grounded in real-world exploit data rather than hypothetical threat assumptions.

\begin{table*}[!thpb]
\renewcommand{\arraystretch}{1.3}
\caption{List of Vulnerabilities with EPSS scores (C: Confidentiality, I: Integrity, A: Availability)}
\label{tab:epss}
\centering
\begin{tabular}{p{3cm} c p{4cm} c c}
\hline
\textbf{Attack Type} & \textbf{CVE} & \textbf{Description} & \textbf{EPSS Score} & \textbf{Impact} \\
\hline
Authentication Bypass & CVE-2020-15506   & Bypass authentication mechanisms  & 0.007  & C \\
Unauthorised Access  & CVE-2018-8360    & Unauthorised access to information  & 0.0268 & C \\
Man-in-the-Middle  & CVE-2014-2361    & Intercepting wireless communication via sensor I/O Module  & 0.0008 & C \\
Man-in-the-Middle  & CVE-2014-2379    & Exploit Sensys Networks for traffic alteration  & 0.0027 & C \\
Side-Channel Attack  & CVE-2004-2768    & Creating hard links for elevated privileges  & 0.0004 & I \\
Side-Channel Attack  & CVE-2014-2362    & Predicting project creation times  & 0.0021 & I \\
Side-Channel Attack  & CVE-2014-2360    & Exploiting battery voltage reporting  & 0.0194 & I \\
Remote Code Execution  & CVE-2020-16898   & Exploiting TCP/IP stack flaw  & 0.0076 & I \\
SQL Injection  & CVE-2019-9568   & Executing arbitrary SQL commands  & 0.002  & I \\
Shell Code Execution  & CVE-2019-17059  & Executing arbitrary commands  & 0.0064 & I \\
Privilege Escalation  & CVE-2004-2768   & Elevating user privileges via hard link  & 0.0004 & I \\
Privilege Escalation  & CVE-2021-22555  & Heap memory corruption for privilege escalation  & 0.0026 & I \\
Denial of Service  & CVE-2014-2361   & Exploitation of wireless communication  & 0.0008 & A \\
Denial of Service  & CVE-2019-0701   & Exploiting input validation flaws  & 0.00004 & A \\
Denial of Service  & CVE-2021-35517  & Memory exhaustion via crafted TAR archive  & 0.0176 & A \\
\hline
\end{tabular}

%\begin{tablenotes}
%\item[1] C: Confidentiality.
%\item[2] I: Integrity.
%\item[3] A: Availability.
%\end{tablenotes}
\end{table*}

\subsubsection{EPSS-Based Quantitative Assignment}

A key methodological extension over \cite{abdulhamid2025b} lies in probability modelling. In the previous safety-oriented framework, exponentially distributed failure rates were assigned to basic events to represent stochastic hardware and communication failures. Such assumptions are inappropriate for cyber attacks, as adversarial exploitation does not follow memoryless reliability distributions.

In this study, EPSS scores are assigned directly to BAS nodes, representing the empirical probability of vulnerability exploitation within a defined time horizon. EPSS values incorporate real-world exploit data, threat intelligence, and vulnerability characteristics, thereby providing a data-driven and dynamically informed estimate of attack likelihood.

Formally, for each BAS $BAS_i$ corresponding to vulnerability $CVE_i$, the probability is assigned according to equation \eqref{EPSS}. These probabilities form the quantitative foundation for subsequent gate-by-gate propagation in the SAT.

\subsubsection{Gate-by-Gate Quantification of the Attack Tree}
\label{sec6}

This section presents the deterministic quantification of the SAT derived from the MBSE artefacts. The objective is to compute the probability of the TE, defined as the security-induced failure of the IoT-enabled SIS, by propagating empirically assigned BAS probabilities through the Boolean structure of the tree.

Each BAS corresponds to a CVE-mapped vulnerability, and its prior probability is obtained directly from the EPSS values reported in Table~\ref{tab:epss}. These probabilities represent empirically derived exploit likelihoods and replace the exponentially distributed failure rates used in the safety-focused framework of \cite{abdulhamid2025b}. 

Quantification proceeds bottom-up, from leaf nodes (BASs) to intermediate attack and sub-attack goals, and ultimately to the TE. The SAT contains \textit{AND} and \textit{OR} logic gates. Under the standard conditional independence assumption, OR and AND gates probabilities are computed using equations \eqref{OR} and \eqref{AND}, respectively.

\paragraph{Illustrative Sub-Attack Calculations:}

To demonstrate the gate-level computation, consider two representative sub-attack goals: Man-in-the-Middle (MiM) and Information Disclosure (ID).

\textbf{MiM Sub-Attack (AND Gate):}  
The MiM goal is realised only if both BASs are successfully exploited:
CVE-2014-2361 (wireless communication vulnerability) and 
CVE-2014-2379 (replay transmission vulnerability).  
As these BASs are connected via an \textit{AND} gate, Equation~\eqref{AND} applies:

\[
\begin{split}
Pr(\text{MiM}) &= Pr(\text{CVE-2014-2361}) \times Pr(\text{CVE-2014-2379}) \\
&= 0.0008 \times 0.0027 \\
&= 2.16 \times 10^{-6}
\end{split}
\]

\textbf{ID Sub-Attack (OR Gate):}  
The ID goal can be achieved through either authentication bypass (CVE-2020-15506) or unauthorised access (CVE-2018-8360). These BASs are connected through an \textit{OR} gate; therefore, Equation~\eqref{OR} applies:

\[
\begin{split}
Pr(\text{ID}) &= 1 - \big(1 - Pr(\text{CVE-2020-15506})\big) \times \\
&\big(1 - Pr(\text{CVE-2018-8360})\big) \\
&= 1 - (1 - 0.007)(1 - 0.0268) \\
&\approx 3.38 \times 10^{-2}
\end{split}
\]

These examples illustrate how EPSS-based BAS probabilities are systematically propagated through the SAT according to its logical structure.

\paragraph{Recursive Propagation to the Top Event:}

The TE probability is obtained by recursively applying Equations~\eqref{OR} and~\eqref{AND} at each hierarchical level of the SAT until the root event is reached. Strict adherence to the SAT structure derived from the MBSE transformation is essential, as both the selection of BAS probabilities and the type of logical gate directly influence the resulting system-level estimate. 

Importantly, vulnerability probabilities must be mapped exclusively to their corresponding BAS nodes. Substituting unrelated CVE data, even if thematically similar, would distort the quantitative interpretation of the attack path and alter the computed TE probability.

Table~\ref{tab:agc} summarises the computed probabilities for all attack and sub-attack goals, including those corresponding to the Confidentiality, Integrity, and Availability (CIA) attributes.

The final computed probability of the TE is:

\[
Pr(\text{TE}) = 4.00 \times 10^{-2}
\]

This value represents the estimated likelihood of a successful cyber-induced system failure within the EPSS reference horizon. Interpreted complementarily, the SIS exhibits an estimated security reliability of approximately $96\%$ with respect to the modelled attack surface and associated vulnerabilities.

It is important to emphasise that this value reflects cyber-induced failure only. The overall system dependability would decrease when combined with random hardware failures, communication faults, or other non-malicious reliability contributors, as considered in \cite{abdulhamid2024enhsaf}. %The integration of such heterogeneous failure mechanisms is addressed in the subsequent Bayesian Network mapping.

 \begin{table*}
\renewcommand{\arraystretch}{1.2}
\caption{Quantitative Estimate of Attack and Sub-Attack Goals}
\label{tab:agc}
%\vspace{1cm} % Adjust the value to shift the table down
\centering
\small
\begin{tabular}{c c c c c c}
\hline
\textbf{Attack} & \textbf{Gate} & \textbf{Pr(Attack)} & \textbf{Attack} & \textbf{Gate} & \textbf{Pr(Attack)} \\
\hline
ID & OR & $3.38 \times 10^{-2}$ & MiM & AND & $2.16 \times 10^{-6}$ \\
%\hline
SC & OR & $2.19 \times 10^{-2}$ & PE & OR & $3.0 \times 10^{-3}$ \\
%\hline
CI & OR & $1.59 \times 10^{-2}$ & UA & AND & $7.3 \times 10^{-8}$ \\
%\hline
DCA & OR & $2.19 \times 10^{-2}$ & DIA & AND & $4.77 \times 10^{-5}$ \\
%\hline
DoS & OR & $1.84 \times 10^{-2}$ & TE & OR & $4.00 \times 10^{-2}$\\
\hline
\end{tabular}
\end{table*}

 \subsubsection{Bayesian Network Mapping and Analysis}
\label{sec:bn}

\begin{figure*}[!thpb]
\centering
\includegraphics[scale=0.9]{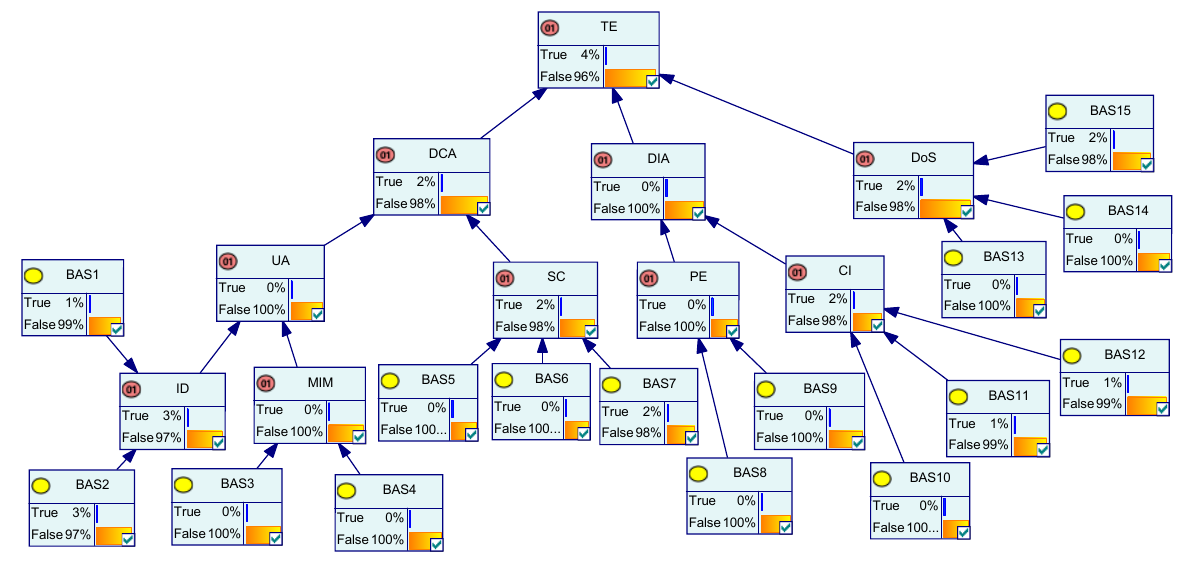}
\caption{ BN Model of the AT of Fig. \ref{description:attacktree}}
\label{description:BN}
\end{figure*}

The System-Level Attack Tree (SAT) was transformed into a BN to enable probabilistic inference beyond deterministic Boolean propagation. The resulting directed acyclic graph, shown in Figure~\ref{description:BN}, preserves the hierarchical structure of the SAT: Basic Attack Steps (BASs) constitute root nodes, intermediate attack and sub-attack goals form internal nodes, and the TE represents the terminal node corresponding to security-induced system failure.

Conditional Probability Tables (CPTs) for intermediate nodes were derived directly from the logical semantics of the SAT gates. Specifically, \textit{AND} and \textit{OR} relationships were encoded using the probabilistic formulations defined in Equations~\eqref{AND} and~\eqref{OR}. Prior probabilities for root nodes (BAS$_1$–BAS$_{16}$) were assigned using the EPSS values reported in Table~\ref{tab:epss}.

\paragraph{Treatment of Repeated CVE Identifiers:}

During the mapping process, two CVE identifiers appeared in multiple attack contexts:

\begin{itemize}
    \item \textbf{CVE-2014-2361}: 
    instantiated once in the Man-in-the-Middle (MiM) attack path (interception of wireless communication via the sensor I/O module), and once in the Denial/Disruption of Service (DoS) path (wireless communication exploitation).
    
    \item \textbf{CVE-2004-2768}: 
    instantiated once in the Side-Channel attack path (hard-link creation leading to information leakage), and once in the Privilege Escalation (PE) path (elevation of user privileges via hard links).
\end{itemize}

Although the CVE identifiers are identical, each occurrence represents a distinct exploitation event embedded in a different architectural and adversarial context. Accordingly, these were modelled as separate BAS nodes in the BN (e.g., BAS$_i$, BAS$_j$), each corresponding to:

\begin{center}
\emph{“successful exploitation of CVE$_k$ within a specific attack path to realise a particular attack goal.”}
\end{center}

These BAS nodes are treated as statistically independent random variables. This assumption reflects that exploitation in one context (e.g., MiM) does not deterministically imply successful exploitation in another context (e.g., DoS), as different access vectors, privileges, or system interfaces may be involved. While correlated exploitation scenarios could be represented through additional dependency arcs, such extensions are outside the scope of the present case study.

\paragraph{Predictive Inference Consistency:}

Forward (predictive) inference in the BN yields a TE probability identical to that obtained through deterministic gate-by-gate propagation, i.e., $Pr(TE)=4.00\times10^{-2}$. This result confirms structural and quantitative consistency between the SAT and its BN representation.

\paragraph{Advanced Bayesian Analyses:}

Beyond predictive estimation, the BN enables analyses not supported by deterministic ATA:

\begin{itemize}
    \item \textbf{Posterior Diagnostic Inference:} 
    Computation of $P(BAS_i \mid TE)$ to identify the most probable contributing vulnerabilities given system compromise.
    
    \item \textbf{Evidence-Based Updating:} 
    Dynamic recalculation of $P(TE \mid \text{evidence})$ under observed attack indicators (e.g., confirmed MiM or authentication bypass events).
    
    \item \textbf{Sensitivity and Criticality Analysis:} 
    Evaluation of vulnerability influence on system failure probability using measures such as Birnbaum Importance to prioritise mitigation efforts.
\end{itemize}

These capabilities transform the static SAT into a bidirectional probabilistic reasoning framework, supporting forensic analysis, real-time risk assessment, and security investment prioritisation.

\subsubsection{Posterior Diagnostic Inference Results}

To evaluate the diagnostic capability of the BN, posterior probabilities of all BAS$_i$ were computed under the condition that the TE has occurred, i.e., $Pr(BAS_i \mid TE=1)$. This reverse inference enables identification of the most probable causal contributors given a confirmed system compromise.

\begin{table*}[!htbp]
\renewcommand{\arraystretch}{1.2}
\caption{Prior and Posterior Probabilities of Basic Attack Steps (Posterior Diagnostic Inference)}
\label{tab:posterior}
\centering
\small
\begin{tabular}{c c c c}
\hline
\textbf{BAS} & \textbf{CVE} & \textbf{Prior Probability} & \textbf{Posterior Probability} \\
\hline
BAS1  & CVE-2020-15506 & $7.0 \times 10^{-3}$  & $7.00035 \times 10^{-3}$ \\
BAS2  & CVE-2018-8360  & $2.68 \times 10^{-2}$ & $2.68013 \times 10^{-2}$ \\
BAS3  & CVE-2014-2361  & $8.0 \times 10^{-4}$  & $8.01745 \times 10^{-4}$ \\
BAS4  & CVE-2014-2379  & $2.7 \times 10^{-3}$  & $2.70174 \times 10^{-3}$ \\
BAS5  & CVE-2004-2768  & $4.0 \times 10^{-4}$  & $1.00202 \times 10^{-2}$ \\
BAS6  & CVE-2014-2362  & $2.1 \times 10^{-3}$  & $5.26063 \times 10^{-2}$ \\
BAS7  & CVE-2014-2360  & $1.94 \times 10^{-2}$ & $4.85982 \times 10^{-1}$ \\
BAS8  & CVE-2004-2768  & $4.0 \times 10^{-4}$  & $5.52735 \times 10^{-4}$ \\
BAS9  & CVE-2021-22555 & $2.6 \times 10^{-3}$  & $3.59278 \times 10^{-3}$ \\
BAS10 & CVE-2019-9568  & $2.0 \times 10^{-3}$  & $2.14196 \times 10^{-3}$ \\
BAS11 & CVE-2019-17059 & $6.4 \times 10^{-3}$  & $6.85428 \times 10^{-3}$ \\
BAS12 & CVE-2020-16898 & $7.6 \times 10^{-3}$  & $8.13946 \times 10^{-3}$ \\
BAS13 & CVE-2014-2361  & $8.0 \times 10^{-4}$  & $2.00405 \times 10^{-2}$ \\
BAS14 & CVE-2019-0701  & $4.0 \times 10^{-5}$  & $1.00202 \times 10^{-3}$ \\
BAS15 & CVE-2021-35517 & $1.76 \times 10^{-2}$ & $4.40891 \times 10^{-1}$ \\
\hline
\end{tabular}
\end{table*}

Table~\ref{tab:posterior} summarises the prior and posterior probabilities of each BAS. Several observations emerge.

First, a subset of BAS nodes exhibits negligible change between prior and posterior probabilities. For example, BAS1 (CVE-2020-15506) and BAS2 (CVE-2018-8360) increase only marginally from 0.007 to 0.00700035 and from 0.0268 to 0.0268013, respectively. This indicates weak causal contribution to the TE within the overall attack structure.

In contrast, other BAS nodes experience substantial posterior amplification. Notably, BAS7 (CVE-2014-2360) increases from 0.0194 to 0.48598, and BAS15 (CVE-2021-35517) rises from 0.0176 to 0.44089. Similarly, BAS6 (CVE-2014-2362) increases from 0.0021 to 0.05261. These large posterior shifts indicate strong causal alignment with attack paths leading to the TE. In probabilistic terms, the occurrence of the TE significantly increases the likelihood that these vulnerabilities were exploited.

A particularly important observation concerns repeated CVE identifiers. CVE-2014-2361 appears as both BAS3 and BAS13 in different attack contexts. While BAS3 shows only marginal posterior increase (0.0008 to 0.0008017), BAS13 exhibits substantial amplification (0.0008 to 0.02004). Similarly, CVE-2004-2768 appears as BAS5 and BAS8, with posterior probabilities of 0.01002 and 0.0005527, respectively. 

These discrepancies confirm that vulnerability impact is strongly path-dependent. Although the CVE identifier is identical, its probabilistic influence differs significantly depending on its architectural embedding and associated attack goal. This validates the modelling decision to represent repeated CVEs as context-specific BAS nodes rather than as a single shared variable.

Overall, the posterior analysis demonstrates that exploitability (as reflected by EPSS priors) does not directly translate into system-level criticality. Certain vulnerabilities with modest prior probabilities become highly probable contributors once the TE is observed. This diagnostic capability, identifying likely root causes after compromise, cannot be achieved using deterministic gate-by-gate ATA and represents a key advantage of the BN framework.

\subsubsection{Birnbaum Importance Ranking and Criticality Analysis}

To further identify the most influential attack steps in the model, a Birnbaum importance analysis was conducted. The Birnbaum importance measure quantifies the sensitivity of the top-level attack probability to changes in the probability of individual basic attack steps. In other words, it evaluates how strongly the likelihood of system compromise depends on the successful exploitation of a specific vulnerability.

%Formally, the Birnbaum importance of a basic attack step $B_i$ with respect to the top-level event $T$ is defined as:

It is defined as the difference between the probability of the TE when a given BAS is assumed to be certainly exploited and the probability of the TE when that BAS is assumed not to occur:

\[
BIM_i = Pr(TE \mid BAS_i=1) - Pr(TE \mid BAS_i=0).
\]

This measure indicates how strongly the exploitation of a specific vulnerability contributes to the occurrence of the system-level attack. Higher BIM values therefore identify vulnerabilities whose presence or absence has the most significant impact on the overall cybersecurity risk.

Using the constructed BN, the Birnbaum importance values were computed for all BAS nodes. Table~\ref{tab:birnbaum_ranking} presents the BIM values for all BAS nodes together with their corresponding rankings. The results reveal a clear differentiation in the relative criticality of vulnerabilities within the system. In particular, BAS7 (CVE-2014-2360) exhibits the highest importance value of 0.979, making it the most critical vulnerability in the model. This indicates that exploitation of this vulnerability substantially increases the probability of the top event and therefore represents the most influential attack pathway in the analysed system. The second most critical vulnerability is BAS15 (CVE-2021-35517) with a BIM value of 0.977, followed by BAS6 (CVE-2014-2362) and BAS13 (CVE-2014-2361), which also demonstrate very high importance values above 0.96.

\begin{table*}[htbp]
\centering
\caption{Birnbaum Importance Measure (BIM) and ranking of Basic Attack Steps.}
\label{tab:birnbaum_ranking}
\begin{tabular}{llll}
\toprule
\textbf{BAS} & \textbf{CVE} & \textbf{BIM} & \textbf{Rank} \\
\midrule
BAS1  & CVE-2020-15506 & 0.0000020190 & 15 \\
BAS2  & CVE-2018-8360  & 0.0000020600 & 14 \\
BAS3  & CVE-2014-2361  & 0.0000871310 & 12 \\
BAS4  & CVE-2014-2379  & 0.0000258160 & 13 \\
BAS5  & CVE-2004-2768  & 0.9604650080 & 5  \\
BAS6  & CVE-2014-2362  & 0.9621012350 & 3  \\
BAS7  & CVE-2014-2360  & 0.9790748750 & 1  \\
BAS8  & CVE-2004-2768  & 0.0152487860 & 8  \\
BAS9  & CVE-2021-22555 & 0.0152824210 & 7  \\
BAS10 & CVE-2019-9568  & 0.0028392100 & 11 \\
BAS11 & CVE-2019-17059 & 0.0028517830 & 10 \\
BAS12 & CVE-2020-16898 & 0.0028552310 & 9  \\
BAS13 & CVE-2014-2361  & 0.9608495020 & 4  \\
BAS14 & CVE-2019-0701  & 0.9601192270 & 6  \\
BAS15 & CVE-2021-35517 & 0.9772809670 & 2  \\
\bottomrule
\end{tabular}
\end{table*}

These results suggest that vulnerabilities associated with sensor behaviour, hardware manipulation, and memory exhaustion attacks play a dominant role in determining the system's susceptibility to cyber-induced failure. Such vulnerabilities lie on critical paths of the attack tree and therefore have a disproportionately large influence on the overall system compromise probability. Consequently, mitigation efforts targeting these vulnerabilities would produce the greatest reduction in system-level cybersecurity risk.

A second tier of vulnerabilities (e.g., BAS8 and BAS9) exhibit moderate importance values (~0.015), indicating contributory but non-dominant roles. Finally, several vulnerabilities exhibit very small BIM values, indicating a negligible influence on the top event probability. For example, BAS1 (CVE-2020-15506) and BAS2 (CVE-2018-8360) have importance values on the order of $10^{-6}$, ranking 15th and 14th, respectively. These vulnerabilities correspond to alternative attack paths that do not significantly affect the probability of system compromise when considered within the overall attack structure. Similarly, BAS4 (CVE-2014-2379) and BAS3 (CVE-2014-2361) show relatively low influence due to their position within AND-gate structures where multiple conditions must be satisfied simultaneously.

Importantly, repeated CVEs again demonstrate context sensitivity. For example, CVE-2004-2768 appears as BAS5 (BIM $\approx$ 0.9605) and BAS8 (BIM $\approx$ 0.0152), showing radically different influence depending on its attack path. Similarly, CVE-2014-2361 appears as BAS3 (BIM $\approx$ 8.7×10$^{-5}$) and BAS13 (BIM $\approx$ 0.9608). These results confirm that vulnerability criticality is governed more by structural position within the attack model than by the CVE identifier itself.

Overall, the Birnbaum importance analysis provides valuable insight into the relative criticality of vulnerabilities, enabling a prioritised approach to cybersecurity risk mitigation. From a defensive perspective, the ranking indicates that mitigation strategies should prioritise vulnerabilities BAS7, BAS15, BAS6, and BAS13, as addressing these vulnerabilities would yield the most significant reduction in the probability of system compromise.

While deterministic gate-by-gate evaluation of the attack tree can estimate the probability of the top event, the Bayesian network enables additional analytical capabilities such as importance ranking and diagnostic reasoning. These capabilities allow security analysts to identify the most influential vulnerabilities and allocate defensive resources more effectively to reduce the likelihood of cyber-induced system failure.

\begin{comment}

\begin{figure*}[!t]
\centering
\begin{tikzpicture}
\begin{axis}[
    width=0.9\linewidth,
    height=7cm,
    ybar,
    bar width=6pt,
    xlabel={Basic Attack Steps (Ranked)},
    ylabel={Birnbaum Importance Measure},
    symbolic x coords={BAS7,BAS15,BAS6,BAS13,BAS5,BAS14,BAS9,BAS8,BAS12,BAS11,BAS10,BAS3,BAS4,BAS2,BAS1},
    xtick=data,
    x tick label style={rotate=45,anchor=east},
    ymin=0,
    grid=major,
]

\addplot coordinates {
(BAS7,0.9790748750)
(BAS15,0.9772809670)
(BAS6,0.9621012350)
(BAS13,0.9608495020)
(BAS5,0.9604650080)
(BAS14,0.9601192270)
(BAS9,0.0152824210)
(BAS8,0.0152487860)
(BAS12,0.0028552310)
(BAS11,0.0028517830)
(BAS10,0.0028392100)
(BAS3,0.0000871310)
(BAS4,0.0000258160)
(BAS2,0.0000020600)
(BAS1,0.0000020190)
};

\end{axis}
\end{tikzpicture}
\caption{Birnbaum importance ranking of Basic Attack Steps derived from the Bayesian Network analysis. Higher BIM values indicate vulnerabilities with greater influence on the probability of the top event.}
\label{fig:BIMranking}
\end{figure*}
\end{comment}
%%%%%%%%%%%%%%%%%%%%%%%%%%%%%%%%%%%%%%%%%%%%%%%%%%%%%%%%%%%%%%%%%%%%%%%%%%%%%%%%%%%%%%%%%%%%

\begin{table*}[t]
\centering
\renewcommand{\arraystretch}{1.25}
\caption{Comparison of Existing Cybersecurity Modelling Approaches with the Proposed Framework}
\label{tab:comp}
\small
\setlength{\tabcolsep}{2pt}
\begin{tabular}{lccccccccc}
\hline
\multirow{2}{*}{\textbf{Studies}} 
& \multicolumn{5}{c}{\textbf{Conceptual Capabilities}} 
& \multicolumn{4}{c}{\textbf{Analytical Capabilities}} \\

\cline{2-10}

& \textbf{Cyber} 
& \textbf{Quantitative} 
& \textbf{Real} 
& \textbf{IoT/ICS} 
& \textbf{MBSE} 
& \textbf{ATA} 
& \textbf{BN} 
& \textbf{Posterior} 
& \textbf{Criticality} \\

& \textbf{Threat} 
& \textbf{Analysis} 
& \textbf{Data} 
& \textbf{Case Study} 
& \textbf{Model} 
& \textbf{Evaluation} 
& \textbf{Inference} 
& \textbf{Reasoning} 
& \textbf{Ranking} \\

\hline

Waqar \textit{et al.} \cite{45} & Y & Y & N & Y & N & Y & N & N & N \\

Mauw \textit{et al.} \cite{mauw2006foundations} & N & N & N & N & N & Y & N & N & N \\

Ten \textit{et al.} \cite{ten2007vulnerability} & Y & Y & N & Y & N & Y & N & N & N \\

Saini \textit{et al.} \cite{saini2008threat} & Y & Y & N & N & N & Y & N & N & N \\

Roy \textit{et al.} \cite{roy2012attack} & Y & Y & N & Y & N & Y & N & N & N \\

Kumar \textit{et al.} \cite{kumar2015quantitative} & Y & N & N & N & N & Y & N & N & N \\

Foster \textit{et al.} \cite{forster2010integration} & N & Y & N & N & N & Y & N & N & N \\

Abdulhamid \textit{et al.} \cite{abdulhamid2025quantitative} & Y & Y & N & Y & N & Y & N & N & N \\

\textbf{Proposed Approach} & \textbf{Y} & \textbf{Y} & \textbf{Y} & \textbf{Y} & \textbf{Y} & \textbf{Y} & \textbf{Y} & \textbf{Y} & \textbf{Y} \\

\hline
\end{tabular}

\begin{tablenotes}
\small
\item Y: Yes; N: No.
\item ATA: Deterministic attack tree evaluation using logical gates.
\item BN: Bayesian Network based probabilistic modelling.
\item Posterior Reasoning: Diagnostic inference when attack evidence is observed.
\item Criticality Ranking: Identification of influential vulnerabilities using importance measures (e.g., Birnbaum Importance).
\item MBSE Model: Use of Model-Based Systems Engineering (e.g., SysML) to derive or support the security model from system architecture.
\item Real Data refers to the use of publicly available vulnerability exploitability datasets such as EPSS and NVD.
\end{tablenotes}

\end{table*}

\section{Discussion and Comparison With Other Approaches}
\label{sec4}
The proposed framework provides a structured methodology for quantitatively analysing cybersecurity threats in IoT-based systems by integrating MBSE, attack tree modelling, and empirical vulnerability exploitability data. In contrast to traditional security analyses that treat system architecture and security models independently, the approach uses SysML models to represent the architectural structure and behavioural interactions of the IoT system, from which component-level attack trees are systematically derived. This model-driven process ensures traceability between system components, associated vulnerabilities, and potential attack propagation paths, enabling security analysis to remain tightly coupled with the underlying system design.

%The proposed framework demonstrates a structured methodology for quantitatively analysing cybersecurity threats in IoT-based systems using real-world vulnerability data. A key distinguishing aspect of the framework is the integration of MBSE with quantitative cybersecurity analysis. Specifically, the architectural structure and behavioural interactions of the IoT system are first represented using SysML models, from which component-level attack trees are systematically derived. This model-driven process ensures that the cybersecurity analysis remains explicitly grounded in the underlying system architecture, enabling traceability between system components, vulnerabilities, and potential attack propagation paths.

%By combining MBSE-based system modelling with attack tree analysis and exploitability data from publicly available vulnerability databases, the approach enables objective estimation of how cyber vulnerabilities may propagate through an IoT architecture and potentially lead to system-level failure conditions. This provides valuable insight into the trustworthiness of IoT-based systems and supports security-aware design decisions during system development.

By incorporating exploitability data obtained from publicly available vulnerability intelligence sources such as the National Vulnerability Database and the EPSS, the framework enables objective estimation of attack likelihoods based on observed exploitation trends. This contrasts with many conventional attack tree analyses that rely on expert judgement, qualitative assessments, or heuristic scoring mechanisms. Consequently, the proposed approach provides a more realistic representation of the threat landscape affecting IoT-based systems.

%Unlike conventional attack tree analyses that rely on expert judgement or subjective scoring, the proposed approach grounds the quantitative analysis in empirical exploitability data obtained from the EPSS. This allows the likelihood of basic attack steps to be estimated based on observed exploitation trends rather than hypothetical attacker profiles. As a result, the analysis provides a more realistic representation of the threat landscape affecting IoT systems.

The mapping of the attack tree model to a Bayesian Network further extends the analytical capabilities of the framework. While deterministic attack tree evaluation allows the computation of the probability of the top event, the Bayesian formulation enables additional probabilistic reasoning. In particular, posterior inference supports diagnostic analysis when evidence of system compromise is observed, while importance measures such as Birnbaum importance allow vulnerabilities to be ranked according to their contribution to system-level failure. These analyses provide actionable insights for prioritising mitigation strategies in complex IoT environments where numerous vulnerabilities may coexist.

To position the proposed framework within the broader landscape of cybersecurity modelling approaches, Table~\ref{tab:comp} compares representative studies employing attack tree or related security modelling techniques with respect to both conceptual modelling scope and analytical capabilities. As shown, existing approaches, including those by Waqar \textit{et al.} \cite{45}, Mauw \textit{et al.} \cite{mauw2006foundations}, Ten \textit{et al.} \cite{ten2007vulnerability}, Saini \textit{et al.} \cite{saini2008threat}, Roy \textit{et al.} \cite{roy2012attack}, Kumar \textit{et al.} \cite{kumar2015quantitative}, Foster \textit{et al.} \cite{forster2010integration}, and Abdulhamid \textit{et al.} \cite{abdulhamid2025quantitative}, primarily rely on deterministic attack tree structures or qualitative reasoning and do not incorporate architectural modelling, real exploitability datasets, or probabilistic inference mechanisms.

In contrast, the proposed framework combines MBSE-based architectural modelling with data-driven quantitative analysis and probabilistic reasoning. This integration enables architecture-aware security modelling, objective estimation of exploitability likelihoods, and systematic identification of critical vulnerabilities, thereby providing stronger analytical support for security-aware system design and cybersecurity risk assessment in IoT-based environments.
\section{Conclusion and Future Works}
\label{sec5}
%This article presents a study that analyses the cyber-attack-induced failure behaviour of an IoT-based environment using a quantitative approach. The study uses a widely used attack tree framework to develop a quantitative model of IoT security analysis that showcases various attack and sub-attack goals. The model demonstrates how these goals can propagate based on identified vulnerabilities to cause a system malfunction.

This article presented a quantitative framework for analysing cybersecurity-induced failure behaviour in IoT-based environments by integrating MBSE, attack tree modelling, and empirical vulnerability exploitability data. The approach uses SysML-based architectural models to derive attack trees that represent potential attack propagation paths within the system. By incorporating exploitability information from vulnerability intelligence sources such as the NVD and the EPSS, the framework enables quantitative estimation of the likelihood of cyber attacks leading to system-level compromise.

To enhance analytical capabilities, the attack tree model was mapped to a Bayesian Network representation. This probabilistic formulation supports advanced reasoning techniques beyond deterministic attack tree evaluation, including posterior diagnostic inference and vulnerability criticality analysis using Birnbaum importance measures. These capabilities allow analysts to identify vulnerabilities that contribute most significantly to system compromise and provide a principled basis for prioritising security mitigation strategies.

%The study pioneers a new paradigm by correlating TE in the ATA model with the actual IoT system's vulnerabilities and supporting it with publicly available vulnerability databases. By utilising modular decomposition and deconstructing the attack tree into specific attack objectives and sub-objectives to pinpoint vulnerabilities, the study helps in utilising data on the exploitation of vulnerabilities in IoT systems to quantify the system's dependability. This innovative approach offers an objective means of evaluating vulnerabilities and their associated risks.

%The study considers the IoT system, its components, operations, protocols, and enabling technologies in IoT design and their vulnerabilities at an abstract level. Therefore, in the future, it is worthwhile to apply the proposed approach to a larger and more complex practical system to perform a rigorous analysis to evaluate the approach's scalability and effectiveness.  Currently, all calculations are based on publicly available data from the National Vulnerability Database. To estimate the complete IoT system's dependability, future research will incorporate other factors, such as random hardware failure and conflicts between co-located systems, and consider a co-analysis of safety and security using a model-based engineering approach.

Overall, the proposed framework demonstrates how architecture-driven modelling, empirical vulnerability intelligence, and probabilistic analysis can be combined to support rigorous cybersecurity risk assessment for IoT-based systems. By linking security analysis directly to system architecture through MBSE, the approach facilitates traceable and data-driven evaluation of cyber risks during system design and analysis.

Despite these advantages, several challenges remain. Although the framework adopts an MBSE-oriented modelling structure, the current implementation still involves substantial manual effort in constructing attack trees, mapping vulnerabilities to system components, and configuring probabilistic models. This may limit scalability when analysing large or highly complex systems.

Future research will therefore focus on developing automated tool support within MBSE environments to streamline model construction, vulnerability mapping, and probabilistic analysis. Integration with SysML-based modelling platforms would enable automated generation of attack models from architectural descriptions and support continuous security analysis throughout the system development lifecycle.

Another challenge arises from the evolving nature of vulnerability intelligence sources such as the NVD and EPSS. Since exploitability scores and vulnerability information change over time as new threats emerge, cybersecurity models must be periodically updated to maintain their validity. Future work will therefore explore dynamic modelling mechanisms that enable automated retrieval of updated vulnerability data and re-computation of probabilistic risk assessments.

Finally, future studies will extend the framework to larger and more complex industrial IoT environments and investigate its integration with broader dependability analysis techniques that consider both safety and security aspects of cyber-physical systems.

%Additionally, a dynamic ATA approach that models time and function-dependent attack behaviour is crucial and will be considered since cybersecurity threats are dynamic. The study proposes mapping the tree to a higher statistical modelling framework, such as a Bayesian network, to overcome the challenges posed by the static nature of the gate relationship. Developing a BN model from the attack tree would yield valuable quantitative attributes for the modelled system. We are currently investigating some of the areas mentioned to provide a comprehensive analysis that objectively co-evaluates security and the other attributes of the dependability of IoT-based environments. This will enable us to achieve the scalability of the proposed approach.

%Furthermore, with the integration of IoT, UAVs, AI, and ML in smart agriculture, there is a pressing need for in-depth research to fully comprehend the complexities of these innovations and their repercussions on the process of security quantification. While certain solutions offer enhanced functionalities and convenience, it is imperative to devise advanced security analysis approaches to effectively counter the rising attack vectors stemming from the increased heterogeneity of diverse systems and algorithms. Research efforts in this area will not only bolster confidence in these innovative solutions but also pave the way for further advancements in the realm of smart agriculture.
%%===========================================================================================%%
\section*{Abbreviations}
% \abbreviations{Abbreviations}
{
The following abbreviations are used in this manuscript:\\
\noindent 
\begin{tabular}{@{}ll}
ACT & Attack Countermeasure Trees \\
ATA & Attack Tree Analysis \\
BAS & Basic Attack Step \\
CIA & Confidentiality, Integrity, and Availability \\
CPS & Cyber-Physical Systems \\
CTMC & Continuous Time Markov Chain \\
CVE & Common Vulnerability Exposure \\
CVSS & Common Vulnerability Scoring System \\
DoS & Denial of Service \\
EPSS & Exploit Prediction Scoring System \\
FT & Fault Tree \\
GPS & Global Positioning System \\
IoMT & Internet-of-Medical Things \\
IoT & Internet of Things \\
MBSE & Model-Based Systems Engineering \\
MiM & Man-in-the-Middle \\
NFP & Non-Functional Properties \\
NVD & National Vulnerability Database \\
QADT & Quantitative Attack-Defence Trees \\
RBD & Reliability Block Diagram \\
RFID & Radio Frequency Identification \\
SIS & Smart Irrigation System \\
SysML/UML & System or Unified Modelling Languages \\
TE & Top Event \\
\end{tabular}

}
%\section*{Declarations}

%\begin{itemize}
%\item Funding: No funding was received for conducting this study.
 %\item Conflict of interest: There is no conflict of interest.
%\end{itemize}

%\begin{acknowledgements}
%\end{acknowledgements}

\bibliographystyle{spmpsci}      % mathematics and physical sciences

\bibliography{sn-bibliography}
%\begin{thebibliography}{}
% etc
%\end{thebibliography}

\end{document}